\documentclass[%
reprint,
superscriptaddress,
groupedaddress,
amsmath,amssymb,
aps,
pra
]{revtex4-1}
\usepackage{graphicx}
\usepackage{dcolumn}
\usepackage{bm}
\usepackage{physics}
\usepackage{color}
\RequirePackage[colorlinks=true,linkcolor=blue,urlcolor=blue,hyperfootnotes=false,citecolor=black]{hyperref}

\begin{document}
\title{Flux-mediated optomechanics with a transmon qubit in the single-photon ultrastrong-coupling regime}

\author{Marios Kounalakis}
\author{Yaroslav M. Blanter}
\author{Gary A. Steele}
\affiliation{Kavli Institute of Nanoscience, Delft University of Technology, 2628 CJ Delft, The Netherlands}

\date{\today}

\begin{abstract}
We propose a scheme for controlling a radio-frequency mechanical resonator at the quantum level using a superconducting qubit.
The mechanical part of the circuit consists of a suspended micrometer-long beam that is embedded in the loop of a superconducting quantum interference device (SQUID) and is connected in parallel to a transmon qubit.
Using realistic parameters from recent experiments with similar devices, we show that this configuration can enable a tuneable optomechanical interaction in the single-photon ultrastrong-coupling regime, where the radiation-pressure coupling strength is larger than both the transmon decay rate and the mechanical frequency.
We investigate the dynamics of the driven system for a range of coupling strengths and find an optimum regime for ground-state cooling, consistent with previous theoretical investigations considering linear cavities.
Furthermore, we numerically demonstrate a protocol for generating hybrid discrete- and continuous-variable entanglement as well as mechanical Schr{\"o}dinger cat states, which can be realised within the current state of the art.
Our results demonstrate the possibility of controlling the mechanical motion of massive objects using superconducting qubits at the single-photon level and could enable applications in hybrid quantum technologies as well as fundamental tests of quantum mechanics.
\end{abstract}
\maketitle
\section{Introduction}
The rapid progress in the field of cavity optomechanics and electromechanics over the last decade has enabled the study of massive micro- and nano-mechanical objects in the quantum regime, paving the way for several technological applications as well as fundamental tests of quantum mechanics~\cite{aspelmeyer2014cavity, rabl2010quantum, marshall2003towards}.
Important advances include ground-state cooling of mechanical resonators~\cite{chan2011laser, teufel2011sideband}, ponderomotive squeezing~\cite{brooks2012non, safavi2013squeezed, purdy2013strong}, coherent state transfer~\cite{palomaki2013entangling}, as well as preparation of quantum states~\cite{reed2017faithful, Hong2017handbury}.
Such optomechanical setups consist of a mechanical drum or beam resonator that is parametrically coupled to a higher-frequency optical or microwave cavity via radiation pressure. 
Typically the coupling $g_0$ is lower than the decay rate of the cavity $\kappa$, limiting the ability to manipulate the mechanical element at single-photon levels.
A strong linearised interaction is effectively achieved by driving the cavity with thousands or even millions of photons, which however leads to unresolved heating issues~\cite{teufel2011sideband, meenehan2014silicon, yuan2015large} and makes it difficult to couple to artificial atoms working in the single-photon regime.

Growing efforts in the field are focusing on reaching the single-photon strong-coupling regime, $g_0>\kappa$, which holds great promise for high-fidelity mechanical state preparation~\cite{aspelmeyer2014cavity}.
An even more intriguing prospect is the possibility of reaching the \emph{single-photon ultrastrong-coupling} regime, where $g_0$ additionally approaches or even exceeds the mechanical frequency $\omega_\text{M}$.
Note that the electromagnetic mode frequency is typically orders of magnitude larger than $g_0$ and $\omega_\text{M}$, corresponding to an ultrastrong-coupling regime of the radiation-pressure interaction in the strong dispersive limit.
This nonlinear regime of optomechanics offers rich opportunities for exploring photon blockade phenomena and generating non-classical mechanical states~\cite{rabl2011photon, nunnenkamp2011single, qian2012quantum, garziano2015single, liao2016macroscopic}.

A promising playground for enhancing the single-photon coupling is flux-mediated optomechanics, in which a vibrating mechanical element parametrically modulates the inductance of a LC microwave cavity.
This can be realised by integrating a mechanical beam into the arms of a superconducting quantum interference device (SQUID), which offers plenty of opportunities for reaching stronger couplings and observing coherent quantum phenomena in Josephson quantum circuits as well as achieving quantum-limited displacement detection~\cite{zhou2006nonlinear, buks2006decoherence, blencowe2007quantum, nation2008quantum, etaki2008motion, nation2013nonclassical}.
This scheme is also particularly promising for coupling mechanical beams to superconducting microwave resonators and qubits in the single-photon ultrastrong-coupling regime~\cite{rimberg2014cavity, nation2016ultrastrong, shevchuk2017strong, khosla2018displacemon}.

A recent experiment has confirmed the viability of this approach using linear SQUID cavities~\cite{rodrigues2019coupling}, however, reaching the single-photon strong-coupling still remains a challenge.
One limitation of this setup was the suppression of the flux susceptibility, which is proportional to the optomechanical coupling, due to the geometric inductance being a considerable fraction of the total inductance of the linear SQUID.
Another limitation of this scheme is related to the fact that the optomechanical coupling is maximised when the applied flux through the SQUID is close to a half-integer flux quantum, which is where the cavity frequency becomes zero.
Moreover, even in the case of ultrastrong coupling between a mechanical resonator and a cavity, the degree of controllability and the range of states that can be created can be limited in the absence of a strong non-linear element, such as a qubit.

Here we show that it is possible to circumvent the issues associated with achieving strong and ultrastrong optomechanical coupling using the approach of Ref.~\cite{rodrigues2019coupling} and demonstrate protocols for controllably cooling down the mechanical mode as well as creating macroscopic mechanical superpositions.
This can be achieved with a modified circuit where the electromagnetic mode is realised using a superconducting transmon qubit~\cite{koch2007charge}.
The two modes are coupled via a flux-mediated radiation-pressure interaction that is tuneable and can reach the single-photon ultrastrong-coupling regime.
Using realistic experimental parameters obtained from recent experiments, we investigate the possibility of cooling the resonator via sideband driving of the qubit and find that ground-state cooling is possible with a fraction of a driving photon, circumventing the issues associated with strong driving and qubits.
Furthermore, we devise an experimentally feasible protocol for creating hybrid Bell-cat entanglement and mechanical Schr\"{o}dinger cat states using flux-pulsing and qubit operations, enabled by the ability to tune the coupling fast and independently of the qubit frequency.
Our results pave the way for the successful on-chip integration of mechanical elements with state-of-the-art transmon-based processors and the manipulation of mechanical motion at single-photon levels, enabling technological applications and fundamental studies of quantum theory.

\section{Electromechanical system}
\begin{figure}[t]
  \begin{center}
  \includegraphics[width=1\linewidth]{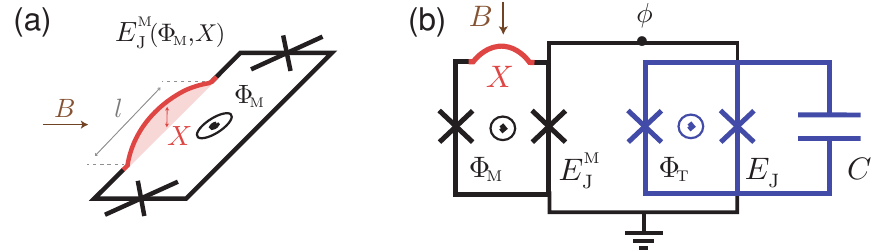}
  \end{center}
  \caption{
  {\bf Proposed circuit architecture.} 
  (a) Schematic representation of the mechanical SQUID comprising a suspended beam that is embedded in the arms of a SQUID loop and can mechanically oscillate out of plane. 
  Upon application of an in-plane magnetic field any mechanical displacement leads to a pick-up flux in the SQUID, resulting in a motion-dependent Josephson inductance.
  (b) The proposed circuit incorporates a flux-tuneable transmon qubit connected in parallel to the mechanical SQUID, leading to a tuneable radiation-pressure coupling between the mechanical resonator and the qubit.
  }
  \label{fig:circuit_RP}
\end{figure}

The mechanical part of the circuit comprises a suspended beam embedded in a SQUID loop that can oscillate out of plane, as schematically depicted in Fig.~\ref{fig:circuit_RP}(a).
Upon application of an in-plane magnetic field ($B$) the loop picks up a flux due to the beam displacement~($X$) which results in a flux-mediated optomechanical interaction between the SQUID cavity and the mechanical oscillator, as recently realised in Ref.~\cite{rodrigues2019coupling}.
Here, we extend this setup by directly connecting it to a superconducting transmon qubit~\cite{koch2007charge}, formed by a second SQUID in parallel to a capacitor $C$.
We refer to the two SQUIDs as the \emph{transmon} and the \emph{mechanical} SQUID, which can be tuned independently by applying locally out-of-plane flux biases $\Phi_\text{T}$ and $\Phi_\text{M}$, respectively.
In an experimental scenario this could be realised via dedicated on-chip flux lines; see e.g. Ref.~\cite{kounalakis2018tuneable} for a realisation in a similar architecture.
The corresponding Josephson energies are given by $E_\text{J}=E_{\text{J,max}}|\cos(\pi\Phi_\text{T}/\Phi_0)|$ and $E_\text{J}^\text{M}(X)=E_{\text{J,max}}^\text{M} |\cos\left(\pi(\Phi_\text{M}+\beta_0 B l X)/\Phi_0)\right|$, where $\Phi_0=h/2e$ is the flux quantum, $l$ is the beam length, and $\beta_0$ is a geometric factor depending on the mechanical mode shape~\cite{etaki2008motion, rodrigues2019coupling}.
The above expression for $E_\text{J}^\text{M}$ is valid for a symmetric SQUID, i.e. when the two junctions are identical.
Including a finite asymmetry $a_\text{J}$ and assuming $\Phi_\text{M}, \Phi_0 \gg \beta_0 B l X$ (see the Appendix), we have
\begin{align}
E_\text{J}^\text{M}~\simeq~E_{\text{J,max}}^\text{M} \left[c_\text{J}\cos(\pi\Phi_\text{M}/\Phi_0) - s_\text{J}\sin(\pi\Phi_\text{M}/\Phi_0) \alpha X\right],
\label{eq:EJx}
\end{align}
where $c_\text{J}=\sqrt{1+a_\text{J}^2\tan{(\pi\Phi_\text{M}/\Phi_0)}}$, $s_\text{J}=(1-a_\text{J}^2)/c_\text{J}$ are correction factors due to the SQUID asymmetry and we have defined $\alpha~\dot{=}~\pi\beta_0 B l/\Phi_0$.

The electromechanical system is described by the Hamiltonian
\begin{align}
&\hat{H}~=~\hat{H}_0 + \hat{H}_\text{int},\\
&\hat{H}_0~=~\hbar\omega_\text{M}~\hat{b}^\dagger\hat{b}+\hbar\omega_\text{T}~\hat{c}^\dagger\hat{c}-\frac{E_\text{C}}{2}~\hat{c}^\dagger\hat{c}^\dagger\hat{c}\hat{c},\\
&\hat{H}_\text{int}~=~ \hbar g_0~\hat{c}^\dagger \hat{c}(\hat{b}+\hat{b}^\dagger)+\hat{H}_\text{int}^\prime,
\label{eq:Ham_quantum}
\end{align}
where $\hat{b}^{(\dagger)}$ and $\hat{c}^{(\dagger)}$ are bosonic operators describing the annihilation (creation) of phonons and qubit excitations, respectively.
The effective transmon frequency is given by $\omega_\text{T}=\left(\sqrt{8\widetilde{E}_\text{J}E_\text{C}}-E_\text{C}\right)/\hbar$, where $\widetilde{E}_\text{J}~=~E_\text{J}+E_{\text{J,max}}^\text{M}c_\text{J}|\cos(\pi\Phi_\text{M}/\Phi_0)|$ is the modified transmon Josephson energy due to the mechanical SQUID and $E_\text{C}~=~{e^2}/{2C}$ is its charging energy.
A detailed derivation of the circuit Hamiltonian is presented in the Appendix.

The qubit is predominantly coupled to the beam via the radiation-pressure interaction, described by the first term in Eq.~(\ref{eq:Ham_quantum}), with single-photon coupling strength
\begin{align}
g_0~&=~\frac{\partial \omega_\text{T}}{\partial \Phi_\text{M}}\frac{\partial \Phi_\text{M}}{\partial X}X_\text{ZPF}\nonumber\\
&=~-\frac{\alpha Z}{2\phi_0^2} s_\text{J}E_{\text{J,max}}^\text{M}\sin(\pi\Phi_\text{M}/\Phi_0) X_\text{ZPF},
\label{eq:g_0}
\end{align}
where $X_\text{ZPF}$ is the zero-point mechanical motion, $Z~=~\frac{\hbar}{e^2}\sqrt{E_\text{C}/2\widetilde{E}_{\text{J}}}$ is the transmon impedance and $\phi_0=\Phi_0/2\pi$ is the reduced flux quantum.

The second term in Eq.~(\ref{eq:Ham_quantum}) describes higher-order contributions to the interaction Hamiltonian (see the Appendix for details)
\begin{equation}
\hat{H}_\text{int}^\prime~=~\hbar g_0^{\prime}~\hat{c}^\dagger \hat{c}^\dagger \hat{c}\hat{c}(\hat{b}+\hat{b}^\dagger)+\hbar g_0^{\prime\prime}~\hat{c}^\dagger \hat{c}(\hat{b}+\hat{b}^\dagger)^2.
\label{eq:Ham_nonlinear}
\end{equation}
The first part is a non-linear correction to the interaction, stemming from the transmon anharmonicity, with coupling strength $g_0^{\prime}=\alpha \hbar Z^2 s_\text{J}E_{\text{J,max}}^\text{M}\sin(\pi\Phi_\text{M}/\Phi_0) X_\text{ZPF}/(16\phi_0^4)$.
Although this term does not impact the dynamics at single-photon levels, it contributes to the radiation-pressure coupling as $g_0\rightarrow g_0+ 2 g_0^{\prime}$.
The second part stems from a higher-order expansion of $E_{\text{J}}^\text{M}$ to $\mathcal{O}[X^2]$, resulting in a coupling strength $g_0^{\prime\prime}~=~\frac{\alpha^2Zs_\text{J}}{4\phi_0^2c_\text{J}}E_{\text{J,max}}^\text{M}\tan(\pi\Phi_\text{b}/\Phi_0)\sin(\pi\Phi_\text{b}/\Phi_0)X_\text{ZPF}^2$ which is three orders of magnitude smaller than $g_0$ for the parameters considered here.
For the sake of completeness we include all terms of $\hat{H}_\text{int}^\prime$ in the simulations, which however lead to negligible effects on the system dynamics.

\begin{figure}[t]
  \begin{center}
  \includegraphics[width=1\linewidth]{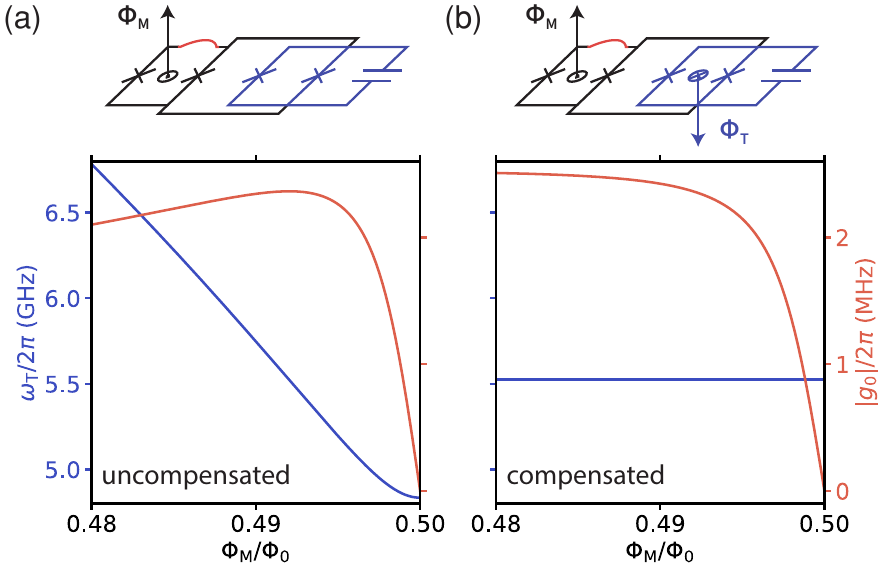}
  \end{center}
  \caption{
  {\bf Tuneable radiation-pressure coupling.} 
  (a) The orange curve corresponds to the single-photon optomechanical coupling as a function of the flux bias on the mechanical SQUID ($\Phi_\text{M}$), while the blue curve depicts the corresponding transmon frequency dependence.
  The coupling becomes zero at $\Phi_\text{M}/\Phi_0=0.5$ as a result of a finite SQUID asymmetry.
  (b) Same plot in the case where an additional flux $\Phi_\text{T}$ is applied on the transmon SQUID (tuning $E_\text{J}$ from 3 to 10 GHz), such that the qubit frequency remains constant while the coupling is tuned.}
  \label{fig:couplings_RP}
\end{figure}

The dependence of the radiation-pressure coupling strength $g_0$, as well as that of the qubit frequency, on the flux bias $\Phi_\text{M}$ is plotted in Fig.~\ref{fig:couplings_RP}(a), for the parameters shown in Table~\ref{tab:Parameters_RP}.
The coupling is maximised at the point where the slope of the qubit frequency ${\partial \omega_\text{T}}/{\partial \Phi_\text{M}}$ is maximum, close to a half-integer flux quantum.
Note that the coupling becomes exactly zero at half-integer flux quanta, as a result of the finite asymmetry of the SQUID [$s_\text{J}$ factor in Eq.~(\ref{eq:g_0})], which is here chosen to be $a_\text{J}=0.01$, reflecting a $2\%$ fabrication error in junction targeting.
Notably, the Josephson inductance of each junction in the SQUID ($L_\text{J}=2\phi_0^2/E_{\text{J,max}}^\text{M}$) is chosen to be much smaller than its expected geometric inductance ($L_\text{g}\sim300$~pH), such that the screening parameter $\beta_\text{L}~=~L_\text{g}/(\pi L_\text{J})\sim0.06$~\cite{clarke2004squid} is negligible and does not limit the achievable coupling strengths as in Ref.~\cite{rodrigues2019coupling}.

Another comparative advantage of this proposal is the additional flux-bias degree of freedom provided by the transmon SQUID.
More specifically, in implementations using a single SQUID, the frequency of the qubit (or SQUID cavity) becomes zero at the point of maximum coupling $\Phi_\text{M}\simeq\Phi_0/2$, as shown in Fig.~\ref{fig:couplings_RP}(a).
Using a second SQUID, however, entirely circumvents this issue as the minimum qubit frequency is set by $E_\text{J}$ and can be non-zero even at $\Phi_\text{M}=\Phi_0/2$.
Most importantly, it allows one to turn the optomechanical coupling on and off while keeping the qubit frequency constant by appropriately adjusting $\Phi_\text{T}$, as depicted in Fig.~\ref{fig:couplings_RP}(b).
This can also ensure that the qubit remains in the transmon regime $E_\text{J}\gg E_\text{C}$, where it is insensitive to charge noise~\cite{koch2007charge}, for the entire coupling range. 

We model the dynamical evolution of the system, using the Lindblad master equation 
\begin{align}
\dot{\rho}~=~&\frac{i}{\hbar}[\rho,\hat{H}]+(n_\text{th}+1)\gamma_m\mathcal{L}[\hat{b}]\rho+n_\text{th}\gamma_m\mathcal{L}[\hat{b}^\dagger]\rho\nonumber\\
&+\frac{(n_\text{th}^\text{T}+1)}{T_1}\mathcal{L}[\hat{c}]\rho+\frac{n_\text{th}^\text{T}}{T_1}\mathcal{L}[\hat{c}^\dagger]\rho+\frac{1}{T_2}\mathcal{L}[\hat{c}^\dagger\hat{c}]\rho,
\label{eq:Lindblad}
\end{align}
where $\mathcal{L}[\hat{o}]\rho=(2\hat{o}\rho\hat{o}^\dagger-\hat{o}^\dagger\hat{o}\rho-\rho\hat{o}^\dagger\hat{o})/2$ are superoperators describing each dissipation process, and $n_\text{th}=1/[\exp(\hbar\omega_m/(k_\text{M}T))-1]$ is the thermal phonon number at temperature $T$.
We use the solver package provided by QuTiP~\cite{johansson2012qutip}, including realistic dissipation rates.
More specifically, we consider qubit decay and dephasing times $T_1=T_2=10~\mu$s which are consistent with measured values in a similar tuneable coupling transmon architecture~\cite{kounalakis2018tuneable} and with transmons operating in 10~mT magnetic fields~\cite{schneider2019transmon}.
We additionally include a thermal transmon occupation $n_\text{th}^\text{T}=5\%$ (effective temperature of 90~mK), corresponding to realistic experimental conditions~\cite{riste2013millisecond, gely2019observation}.
The coupling of the mechanical mode to the environment is determined by $\gamma_m=\omega_m/Q$, where the quality factor $Q=10^6$ is chosen in agreement with experimental observations in recently fabricated SQUID-embedded beams~\cite{rodrigues2019coupling}.

\begin{table}[h]
\begin{tabular}{l c }
\hline
\cline{1-2}
Parameter\ \ \ \ \ \ \ \ \  & Value  \\ 
\hline
\hline
$\omega_\text{M}/(2\pi)$   & 1 MHz \\
$\omega_\text{T}/(2\pi)$   & 5.53 GHz\\
$|g_0|/(2\pi)$   & $\leq2.4$ MHz\\
$E_{\text{J,max}}^\text{M}/h$  & 200 GHz \\  
$E_{\text{J}}/h$  & 3-10 GHz \\  
$E_{\text{C}_i}/h$     & 280 MHz  \\
$B$ & 10 mT \\
$\Phi_\text{M}/\Phi_0$    & 0.49-0.5  \\
$l~$ & 147 $\mu$m \\
$\beta_0~$ & 1 \\
$n_\text{th}$  & $\sim 200$~(10~mK)  \\  
$T_1,~T_2$  & 10  $\mu$s \\
$Q_\text{M}$  & $10^6$ \\
\hline
\cline{1-2}
\end{tabular}
\caption{Parameter set used in the numerical simulations.}
 \label{tab:Parameters_RP}
\end{table}

\section{Ground-state cooling}
Manipulating the mechanical oscillator at the quantum level requires the ability to cool it down to its ground state, where thermal effects are suppressed.
In typical optomechanical setups, this is achieved via a red-detuned continuous-wave (CW) tone on the electromagnetic resonator~\cite{chan2011laser, teufel2011sideband}.
This leads to an effective linearised interaction that is used to transfer phonons to the resonator, which eventually decay.
Typically the single-photon coupling is small and thousands of drive photons are required, therefore the success of these schemes relies heavily on the resonator being linear.
CW ground-state cooling via a transmon qubit has the additional disadvantage of the pump power being limited by the critical photon number in Josephson junctions~\cite{lescanne2019observing}.
These issues could be circumvented in a time-domain scheme, by employing an additional qubit and combining tripartite photon-phonon SWAP gates with qubit reset~\cite{kounalakis2019synthesizing}, which is however outside the scope of this study.

\begin{figure}[t]
  \begin{center}
  \includegraphics[width=1\linewidth]{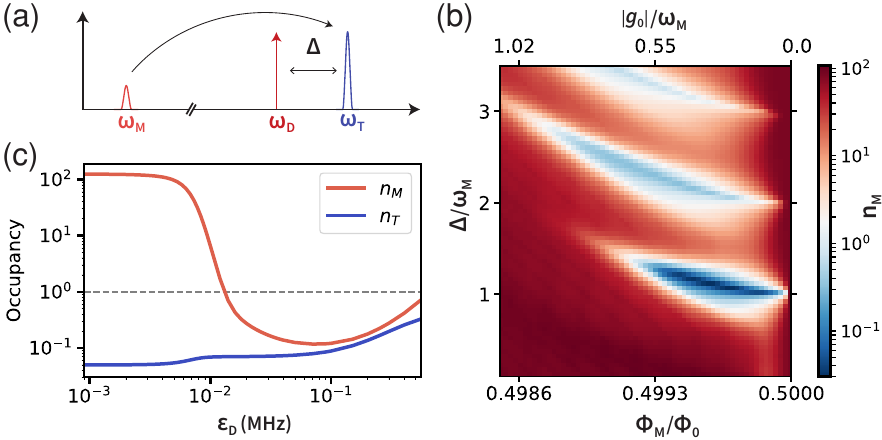}
  \end{center}
  \caption{
  {\bf Ground-state cooling.} 
  (a) Sideband cooling scheme where a red-detuned drive is applied on the qubit.
  (b) Steady-state phonon occupancy as a function of the detuning $\Delta$ and flux bias $\Phi_\text{M}$ corresponding to different ratios of $|g_0|/\omega_\text{M}$ (shown in the top horizontal axis), for a driving amplitude of $\mathcal{E}_\text{D}=70$~kHz.
  (c) Cooling as a function of $\mathcal{E}_\text{D}$ near the optimal condition of $|g_0|\simeq\omega_\text{M}/4$ found in (b).
  We additionally include a $5\%$ thermal transmon occupation, which limits the ground-state cooling to a phonon occupancy of $\sim10\%$.
  }  
  \label{fig:Cooling_RP}
\end{figure}

We investigate the possibility of cooling the beam via sideband driving on the transmon qubit, as depicted in Fig.~\ref{fig:Cooling_RP}(a).
More specifically, we add a driving term $\hat{H}_\text{D}/\hbar~=~ \mathcal{E}_\text{D}(\hat{c}^\dagger e^{-i\omega_\text{D}t} + \hat{c}e^{i\omega_\text{D}t})$ to the system Hamiltonian, where $\mathcal{E}_\text{D}$ and $\omega_\text{D}$ denote the amplitude and frequency of the driving tone, respectively.
We numerically solve the Lindblad equation (\ref{eq:Lindblad}) for the set of parameters listed in Table~\ref{tab:Parameters_RP}.
The choice of the standard master equation is justified by the fact that we employ a qubit and the number of drive photons is always kept well below the single-photon level.
For the sake of completeness, however, our results may also be tested in the future using a dressed-state optomechanical master equation~\cite{hu2015quantum, naseem2018thermodynamic}, suitably modified for qubits, and under the assumption of coloured noise as noted in Refs.~\cite{beaudoin2011dissipation, settineri2018dissipation}.

In Fig.~\ref{fig:Cooling_RP}(b) we plot the steady-state occupation in the mechanical resonator as a function of the detuning $\Delta=\omega_\text{T}-\omega_\text{D}$ and single-photon coupling strength for $\mathcal{E}_\text{D}=70$~kHz.
Note that the number of drive photons $n_\text{D}\simeq(\mathcal{E}_\text{D}/\Delta)^2$ is always well below the single-photon level.
The cooling resonances observed at multiples of $\Delta=\omega_\text{M}-g_0^2/\omega_\text{M}$ are in accordance with predictions for weakly driven optomechanical systems in the single-photon strong-coupling regime~\cite{nunnenkamp2012cooling}.
We find an optimal cooling regime around $|g_0|=\omega_\text{M}/4$, leading to a phonon occupancy of 3\% (assuming a perfectly thermalised transmon $n_\text{th}^{T}=0$).
As expected, ground-state cooling with a qubit becomes impossible in the limit of small coupling $|g_0|\ll\omega_m$, or as $|g_0|\rightarrow\omega_\text{M}$ when the driving tone causes the two modes to hybridise.
In Fig.~\ref{fig:Cooling_RP}(c), we plot the steady-state phonon occupancy as we vary the amplitude of the drive, for the optimal cooling condition found in (b) and including a transmon thermal occupancy of $n_\text{th}^{T}=5\%$, leading to $n_\text{M}^\text{min}\simeq10\%$ at $\mathcal{E}_\text{D}=70$~kHz.

\section{Mechanical cat states}
\begin{figure}[t]
  \begin{center}
  \includegraphics[width=1\linewidth]{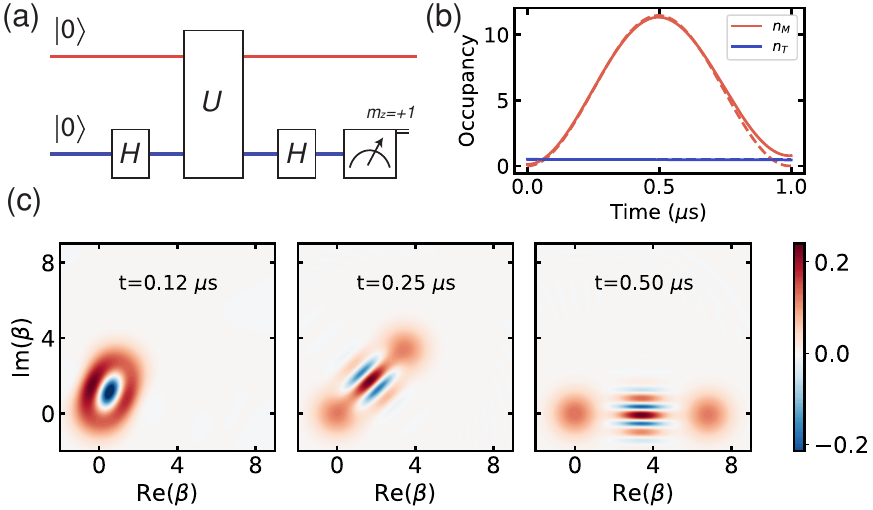}
  \end{center}
  \caption{
  {\bf Generating mechanical cat states.} 
  (a) Description of the protocol.
  In the first step, with the coupling turned off ($\Phi_\text{M}/\Phi_0=0.5$), the qubit is prepared in the superposition state $|+\rangle_\text{T}$ by applying a Hadamard gate.
  The optomechanical coupling is turned on ($\Phi_\text{M}/\Phi_0=0.49$) for a variable time such that the system evolves under the radiation-pressure interaction $U(t)=e^{-ig_0t\hat{c}^\dagger \hat{c}(\hat{b}+\hat{b}^\dagger)}$.
  The coupling is then turned off and a second Hadamard gate is applied on the qubit, followed by a measurement in the computational basis.
  Measuring the qubit in the ground or excited state results in even or odd Schr\"{o}dinger cat states in the mechanical oscillator.
  (b) Evolution of the system excitations after preparing the qubit in a superposition state and turning on the interaction.
  The blue curve corresponds to the qubit excitation number while the orange curve depicts the phonon number evolution for one cycle ($t=1/\omega_\text{M}$), including 0.1 and 0.05 thermal phonon and qubit occupancy, respectively, as obtained from Fig.~\ref{fig:Cooling_RP}(c).
  The dashed curves correspond to the ideal evolution of the system without dissipation.
  (c) Wigner functions of the mechanical resonator at different times following the protocol in (a) and projecting on $|0\rangle_\text{T}$.
  At $t=1/2\omega_\text{M}$ an even cat state is created with $98\%$ ($93\%$) fidelity, starting from an ideal (attainable) ground state.}
  \label{fig:CatState}
\end{figure}

The ultrastrong optomechanical interaction can be used to create arbitrary mechanical states~\cite{garziano2015single}.
In this section we propose an experimentally feasible protocol for high-fidelity preparation of macroscopic mechanical superposition states using the transmon qubit, which is schematically presented in Fig.~\ref{fig:CatState}(a).
The protocol relies on an important feature of the proposed circuit, namely that it allows for fast switching of the optomechanical interaction from zero to ultrastrong coupling and without affecting the qubit frequency.
Practically this can be achieved by applying flux pulses via dedicated on-chip lines, as short as a few nanoseconds, i.e. much shorter than the interaction timescales.

In the first step of the protocol, starting from the ground state $|0\rangle_\text{T}|0\rangle_\text{M}$ and with the coupling off ($\Phi_\text{M}/\Phi_0=0.5$), a Hadamard gate is applied on the qubit, which creates the superposition state $|+\rangle_\text{T}~=~\frac{1}{\sqrt{2}}(|0\rangle+|1\rangle)_\text{T}$.
The second step consists of flux-pulsing the mechanical SQUID to $\Phi_\text{M}/\Phi_0=0.49$ and letting the system evolve for a variable time under the radiation-pressure interaction $U(t)=e^{-ig_0t\hat{c}^\dagger \hat{c}(\hat{b}+\hat{b}^\dagger)}$.
The evolution of excitations in the system after one cycle ($t=1/\omega_\text{M}$) is plotted in Fig.~\ref{fig:CatState}(b), assuming thermal occupancies of $10\%$ and $5\%$ for the beam and the qubit, respectively.
The operation of $U(t)$ results in a coherent displacement on the mechanical resonator depending on the qubit being in the excited state, i.e.
\begin{equation}
U(t)|+\rangle_\text{T}|0\rangle_\text{M}~=~\frac{1}{\sqrt{2}}(|0\rangle_\text{T}|0\rangle_\text{M}+|1\rangle_\text{T}|\beta\rangle_\text{M}),
\label{eq:CVentangled}
\end{equation}
where $|\beta\rangle_\text{M}$ denotes a coherent mechanical state of amplitude $\beta=\sqrt{n_\text{M}}$.

The state created above resembles a hybrid Bell-cat state featuring discrete-continuous variable entanglement~\cite{brune1992manipulation, vlastakis2015characterizing} and can also be written as
\begin{equation}
|\psi\rangle_\text{TM}=\frac{1}{\sqrt{2}}|+\rangle_\text{T}\left(|0\rangle+|\beta\rangle\right)_\text{M}+|-\rangle_\text{T}\left(|0\rangle-|\beta\rangle\right)_\text{M}.
\end{equation}
Turning off the coupling and applying a second Hadamard gate on the qubit, transforms the state into
\begin{equation}
|\psi\rangle_\text{TM}=\frac{1}{2\sqrt{2}}|0\rangle_\text{T}\left(|0\rangle+|\beta\rangle\right)_\text{M}+|1\rangle_\text{T}\left(|0\rangle-|\beta\rangle\right)_\text{M}.
\end{equation}
By performing a projective measurement on the qubit, the beam collapses in a macroscopic superposition $|0\pm\beta\rangle$, depending on whether the qubit is measured in its ground or excited state.
Specifically, this state corresponds to an even or odd Schr\"{o}dinger cat state displaced by $\beta/2$.
In Fig.~\ref{fig:CatState}(c) we plot the evolution of the even cat state after repeating the above protocol and conditioning on $|0\rangle_\text{T}$.
The size of the cat state is maximum at half a cycle and is determined by $\beta_\text{max}=2|g_0|/\omega_\text{M}$.
The fidelity of the prepared state to the ideal Schr\"{o}dinger cat state is $93\%$ and is mainly limited by the finite thermal occupancy of the initial ground state.
Assuming no initial thermal occupancy, we find $98\%$ cat state fidelity, while for ideal evolution without dissipation the fidelity is $99.8\%$.
All higher order interaction terms in Eq.~(\ref{eq:Ham_nonlinear}) are included in the simulations.

\section{Discussion}
In summary, we have analysed a hybrid system involving a superconducting transmon qubit parametrically coupled to a mechanical beam via radiation pressure in the ultrastrong-coupling regime, where the coupling strength exceeds the mechanical frequency at the single-photon level.
We use experimentally feasible parameters, which have been reported in recent experiments combining aluminium beams with SQUIDs~\cite{rodrigues2019coupling}, and small magnetic fields that do not compromise the performance of transmon qubits below $\sim10~\mu$s~\cite{schneider2019transmon}.
We have investigated numerically the possibility of ground-state cooling, by sideband driving below the single-photon level, for a range of achievable coupling strengths.
Additionally, we have devised a proof-of-concept protocol for preparing mechanical Schr\"{o}dinger cat states with high fidelity, benefiting from the tuneable ultrastrong optomechanical interaction.

Our proposed circuit architecture provides a versatile platform for integrating transmon qubits with long-lived mechanical resonators, and may find interesting applications in hybrid quantum technologies~\cite{kurizki2015quantum}.
More specifically, the prepared Bell-cat states are particularly interesting for several quantum computing schemes and error correcting protocols~\cite{cochrane1999macroscopically, jeong2002efficient, leghtas2013hardware, andersen2015hybrid}.
Such macroscopically distinct massive superposition states are also ideally suited for testing fundamental aspects of quantum theory and its relation to gravity~\cite{penrose1996gravity, kleckner2008creating}.
Furthermore, the radiation-pressure interaction can be employed to prepare mechanical Gottesman--Kitaev--Preskill states, which are useful for fault-tolerant error correction schemes~\cite{gottesman2001encoding, weigand2019realizing}.\\

The simulation code supporting the findings of this work is available in Zenodo with identifier 10.5281/zenodo.3776468.

\section*{Acknowledgements}
We thank M.F. Gely, D. Bothner and I.C. Rodrigues for useful discussions.
This work was supported by the Netherlands Organisation for Scientific Research (NWO/OCW), as part of the Frontiers of Nanoscience program and the European Research Council under the European Unions H2020 program [grant number 828826-Quromorphic].


\section*{Appendix: Detailed analysis of the electromechanical system}

\renewcommand{\thesection}{A\arabic{section}}
\renewcommand{\theequation}{A\arabic{equation}}
\renewcommand{\thefigure}{A\arabic{figure}}
\renewcommand{\thetable}{A\arabic{table}}

\subsection{Circuit Hamiltonian}

The Lagrangian describing the electromechanical circuit in Fig.~\ref{fig:circuit_RP}(b) is
\begin{align}
\mathcal{L}~=&~\frac{m\dot{X}^2}{2} -\frac{m\omega_\text{M}^2X^2}{2}+\frac{1}{2}C\dot{\phi}^2  \nonumber\\
&+ [E_\text{J}+E_\text{J}^\text{M}(X)]\cos{\left(\frac{\phi}{\phi_0}\right)},
 \label{eq:Langragian}
\end{align}
where $X,~\phi$ are variables representing the beam displacement and the node flux, respectively, and $\phi_0=\hbar/2e$ is the reduced flux quantum.
$C$ denotes the total capacitance of the transmon and Josephson junctions, which are added in parallel.
Following a Legendre transformation we obtain the system Hamiltonian
\begin{align}
H~=&~ \frac{P^2}{2 m} + \frac{m\omega_\text{M}^2X^2}{2} \nonumber\\
&+\frac{Q^2}{2C}- [ E_\text{J}+E_\text{J}^\text{M}(X)]\cos{\left(\frac{\phi}{\phi_0}\right)},
\label{eq:CircuitHam}
\end{align}
where $\{X,~P\}$ and $\{\phi,~Q\}$ are conjugate variable pairs describing the mechanical and the electrical degrees of freedom, respectively.

The optomechanical coupling between the resonator and the qubit can be determined by analysing the term $E_\text{J}^\text{M}(X)\cos{\left(\frac{\phi}{\phi_0}\right)}$ in the above equation.
The motion-dependent Josephson energy of the mechanical SQUID is given by
\begin{align}
E_\text{J}^\text{M}(\Phi_\text{b}, X)~=~E_{\text{J,max}}^\text{M} [&\cos^2(\pi\Phi_\text{b}/\Phi_0 + \alpha X)\nonumber\\
&+a_\text{J}^2\sin^2(\pi\Phi_\text{b}/\Phi_0+\alpha X)]^{1/2},
\label{eq:EJm_exact}
\end{align}
where $a_\text{J}$ is the SQUID asymmetry.
Following the analysis presented in Ref.~\cite{kounalakis2019synthesizing}, for $\alpha X\ll1,\pi\Phi_\text{b}/\Phi_0$, this expression can be approximated by
\begin{align}
E_\text{J}^\text{M}~\simeq~E_{\text{J,max}}^\text{M}& \left[c_\text{J}\cos(\pi\Phi_\text{b}/\Phi_0) - s_\text{J}\sin(\pi\Phi_\text{b}/\Phi_0) \alpha X\right],
\label{eq:EJm_x}
\end{align}
up to $\mathcal{O}[X]$, where $c_\text{J}~=~\sqrt{1+a_\text{J}^2\tan{(\pi\Phi_\text{M}/\Phi_0)}}$ and $s_\text{J}~=~(1-a_\text{J}^2)/c_\text{J}$.

\subsection{Circuit quantisation and radiation-pressure coupling strength}

The first term in Eq.~(\ref{eq:EJm_x}) results in an effective transmon Josephson energy given by
\begin{equation}
\widetilde{E}_{\text{J}}~=~E_{\text{J}}+E_{\text{J,max}}^\text{M}c_\text{J}\cos(\pi\Phi_\text{M}/\Phi_0),
\end{equation}
which is responsible for the qubit frequency change as a function of $\Phi_\text{M}$, shown in Fig.~\ref{fig:couplings_RP}(a).
The second term, combined with an expansion of the cosine term in Eq.~(\ref{eq:CircuitHam}) up to $\mathcal{O}[\phi^4]$, yields the optomechanical interaction Hamiltonian
\begin{align}
H_\text{int}~=&~-\alpha~E_{\text{J,max}}^\text{M}s_\text{J} \sin(\pi\Phi_\text{M}/\Phi_0) X \left(\frac{\phi^2 }{2\phi_0^2}-\frac{\phi^4 }{24\phi_0^4}\right).
\label{eq:Ham_int}
\end{align}

We can express the interaction Hamiltonian in second quantisation form by promoting all canonical variables to quantum operators
\begin{align}
&\hat{X}~=~X_\text{ZPF}~(\hat{b}+\hat{b}^\dagger),~\hat{P}~=~P_\text{ZPF}~i(\hat{b}^\dagger-\hat{b}),\nonumber\\
&\hat{\phi}~=~\sqrt{\frac{\hbar Z}{2}}~(\hat{c}+\hat{c}^\dagger),~\hat{Q}~=~\sqrt{\frac{\hbar}{2Z}}~i(\hat{c}^\dagger-\hat{c}),
\label{eq:2ndQuant}
\end{align}
where $\hat{b}^{(\dagger)}$, $\hat{c}^{(\dagger)}$ are ladder operators describing the annihilation (creation) of phonons and qubit excitations, respectively, satisfying bosonic commutation relations $[\hat{c},~\hat{c}^\dagger]~=~1$ and $[\hat{b},~\hat{b}^\dagger]~=~1$. 
The zero-point fluctuations in the mechanical displacement and momentum are given by $X_\text{ZPF}~=~\sqrt{{\hbar}/{(2m\omega_\text{M})}}$ and $P_\text{ZPF}~=~\sqrt{{\hbar m\omega_\text{M}}/{2}}$, respectively.
$Z=\frac{\hbar}{e^2}\sqrt{E_\text{C}/2\widetilde{E}_{\text{J}}}$ denotes the transmon impedance, where $E_\text{C}~=~\frac{e^2}{2C}$ is its charging energy, and the qubit frequency is given by $\omega~=~\frac{1}{\hbar}\left(\sqrt{8\widetilde{E}_{\text{J}}E_\text{C}}-E_\text{C}\right)$.

Replacing the classical variables in Eq.~(\ref{eq:Ham_int}) with the quantum operators introduced in Eq.~(\ref{eq:2ndQuant}) we have
\begin{align}
\hat{H}_\text{int}/\hbar
~=g_0 \hat{c}^\dagger \hat{c}(\hat{b}+\hat{b}^\dagger) + g_0^{\prime}\hat{c}^\dagger \hat{c}^\dagger \hat{c}\hat{c}(\hat{b}+\hat{b}^\dagger),
\end{align}
following a rotating wave approximation (RWA) where fast rotating terms $(\hat{c}^{(\dagger)})^n$ are neglected.
The first term describes a radiation-pressure interaction between the qubit and the resonator with coupling strength
\begin{align}
g_0~=~\frac{\alpha Z}{2\phi_0^2} E_{\text{J,max}}^\text{M}\sin(\pi\Phi_\text{M}/\Phi_0) X_\text{ZPF}.
\end{align}
The second term is a higher-order correction to the interaction, stemming from the transmon anharmonicity, with coupling strength
\begin{align}
g_0^{\prime}=\alpha \hbar Z^2 s_\text{J}E_{\text{J,max}}^\text{M}\sin(\pi\Phi_\text{M}/\Phi_0) X_\text{ZPF}/(16\phi_0^4).
\end{align}
This term is included in the simulations although it does not lead to substantial contribution in the system dynamics; however, it leads to a correction of the radiation pressure coupling, $g_0\rightarrow g_0+2g_0^{\prime}$.\\

\subsection{Higher-order interaction terms}
The next-to-leading order correction in the expansion of $E_{\text{J}}^\text{M}(X)$ in Eq.~(\ref{eq:EJm_x}) is given by
\begin{equation}
E_{\text{J}}^\text{M}\{\mathcal{O}[X^2]\}~=~-E_{\text{J,max}}^\text{M}\frac{s_\text{J}\sin^2(\pi\Phi_\text{b}/\Phi_0)}{2c_\text{J}\cos(\pi\Phi_\text{b}/\Phi_0)}\alpha^2X^2.
\end{equation}
This term, combined with a second-order expansion of the cosine in Eq.~(\ref{eq:CircuitHam}), yields the following interaction
\begin{equation}
H_\text{int}^{\{\phi^2X^2\}}~=~E_{\text{J,max}}^\text{M}\frac{s_\text{J}\sin^2(\pi\Phi_\text{b}/\Phi_0)}{2c_\text{J}\cos(\pi\Phi_\text{b}/\Phi_0)}\alpha^2X^2\frac{\phi^2}{2\phi_0},
\end{equation}
which can be written in second quantisation form (following a RWA) as 
\begin{equation}
H_\text{int}^{\{\phi^2X^2\}}~=~g_0^{\prime\prime}\hat{c}^\dagger \hat{c}(\hat{b}+\hat{b}^\dagger)^2.
\end{equation}
The coupling strength of this interaction is given by
\begin{equation}
g_0^{\prime\prime}~=~\frac{\alpha^2Zs_\text{J}}{4\phi_0^2c_\text{J}}E_{\text{J,max}}^\text{M}\tan(\pi\Phi_\text{b}/\Phi_0)\sin(\pi\Phi_\text{b}/\Phi_0)X_\text{ZPF}^2.
\end{equation}
The maximum value of this coupling strength is $g_0^{\prime\prime}~\simeq~5$~kHz around $\Phi_\text{M}/\Phi_0=0.497$ for the parameters considered in this work.
For the sake of completeness we include this interaction in the simulations; however, we do not observe any significant impact on the fidelity of the cooling and quantum state preparation protocols.


\begin{thebibliography}{54}%
\makeatletter
\providecommand \@ifxundefined [1]{%
 \@ifx{#1\undefined}
}%
\providecommand \@ifnum [1]{%
 \ifnum #1\expandafter \@firstoftwo
 \else \expandafter \@secondoftwo
 \fi
}%
\providecommand \@ifx [1]{%
 \ifx #1\expandafter \@firstoftwo
 \else \expandafter \@secondoftwo
 \fi
}%
\providecommand \natexlab [1]{#1}%
\providecommand \enquote  [1]{``#1''}%
\providecommand \bibnamefont  [1]{#1}%
\providecommand \bibfnamefont [1]{#1}%
\providecommand \citenamefont [1]{#1}%
\providecommand \href@noop [0]{\@secondoftwo}%
\providecommand \href [0]{\begingroup \@sanitize@url \@href}%
\providecommand \@href[1]{\@@startlink{#1}\@@href}%
\providecommand \@@href[1]{\endgroup#1\@@endlink}%
\providecommand \@sanitize@url [0]{\catcode `\\12\catcode `\$12\catcode
  `\&12\catcode `\#12\catcode `\^12\catcode `\_12\catcode `\%12\relax}%
\providecommand \@@startlink[1]{}%
\providecommand \@@endlink[0]{}%
\providecommand \url  [0]{\begingroup\@sanitize@url \@url }%
\providecommand \@url [1]{\endgroup\@href {#1}{\urlprefix }}%
\providecommand \urlprefix  [0]{URL }%
\providecommand \Eprint [0]{\href }%
\providecommand \doibase [0]{http://dx.doi.org/}%
\providecommand \selectlanguage [0]{\@gobble}%
\providecommand \bibinfo  [0]{\@secondoftwo}%
\providecommand \bibfield  [0]{\@secondoftwo}%
\providecommand \translation [1]{[#1]}%
\providecommand \BibitemOpen [0]{}%
\providecommand \bibitemStop [0]{}%
\providecommand \bibitemNoStop [0]{.\EOS\space}%
\providecommand \EOS [0]{\spacefactor3000\relax}%
\providecommand \BibitemShut  [1]{\csname bibitem#1\endcsname}%
\let\auto@bib@innerbib\@empty
\bibitem [{\citenamefont {Aspelmeyer}\ \emph {et~al.}(2014)\citenamefont
  {Aspelmeyer}, \citenamefont {Kippenberg},\ and\ \citenamefont
  {Marquardt}}]{aspelmeyer2014cavity}%
  \BibitemOpen
  \bibfield  {author} {\bibinfo {author} {\bibfnamefont {M.}~\bibnamefont
  {Aspelmeyer}}, \bibinfo {author} {\bibfnamefont {T.~J.}\ \bibnamefont
  {Kippenberg}}, \ and\ \bibinfo {author} {\bibfnamefont {F.}~\bibnamefont
  {Marquardt}},\ }\href {\doibase 10.1103/RevModPhys.86.1391} {\bibfield
  {journal} {\bibinfo  {journal} {Rev. Mod. Phys.}\ }\textbf {\bibinfo {volume}
  {86}},\ \bibinfo {pages} {1391} (\bibinfo {year} {2014})}\BibitemShut
  {NoStop}%
\bibitem [{\citenamefont {Rabl}\ \emph {et~al.}(2010)\citenamefont {Rabl},
  \citenamefont {Kolkowitz}, \citenamefont {Koppens}, \citenamefont {Harris},
  \citenamefont {Zoller},\ and\ \citenamefont {Lukin}}]{rabl2010quantum}%
  \BibitemOpen
  \bibfield  {author} {\bibinfo {author} {\bibfnamefont {P.}~\bibnamefont
  {Rabl}}, \bibinfo {author} {\bibfnamefont {S.~J.}\ \bibnamefont {Kolkowitz}},
  \bibinfo {author} {\bibfnamefont {F.}~\bibnamefont {Koppens}}, \bibinfo
  {author} {\bibfnamefont {J.}~\bibnamefont {Harris}}, \bibinfo {author}
  {\bibfnamefont {P.}~\bibnamefont {Zoller}}, \ and\ \bibinfo {author}
  {\bibfnamefont {M.~D.}\ \bibnamefont {Lukin}},\ }\href@noop {} {\bibfield
  {journal} {\bibinfo  {journal} {Nature Physics}\ }\textbf {\bibinfo {volume}
  {6}},\ \bibinfo {pages} {602} (\bibinfo {year} {2010})}\BibitemShut {NoStop}%
\bibitem [{\citenamefont {Marshall}\ \emph {et~al.}(2003)\citenamefont
  {Marshall}, \citenamefont {Simon}, \citenamefont {Penrose},\ and\
  \citenamefont {Bouwmeester}}]{marshall2003towards}%
  \BibitemOpen
  \bibfield  {author} {\bibinfo {author} {\bibfnamefont {W.}~\bibnamefont
  {Marshall}}, \bibinfo {author} {\bibfnamefont {C.}~\bibnamefont {Simon}},
  \bibinfo {author} {\bibfnamefont {R.}~\bibnamefont {Penrose}}, \ and\
  \bibinfo {author} {\bibfnamefont {D.}~\bibnamefont {Bouwmeester}},\ }\href
  {\doibase 10.1103/PhysRevLett.91.130401} {\bibfield  {journal} {\bibinfo
  {journal} {Phys. Rev. Lett.}\ }\textbf {\bibinfo {volume} {91}},\ \bibinfo
  {pages} {130401} (\bibinfo {year} {2003})}\BibitemShut {NoStop}%
\bibitem [{\citenamefont {Chan}\ \emph {et~al.}(2011)\citenamefont {Chan},
  \citenamefont {Alegre}, \citenamefont {Safavi-Naeini}, \citenamefont {Hill},
  \citenamefont {Krause}, \citenamefont {Gr{\"o}blacher}, \citenamefont
  {Aspelmeyer},\ and\ \citenamefont {Painter}}]{chan2011laser}%
  \BibitemOpen
  \bibfield  {author} {\bibinfo {author} {\bibfnamefont {J.}~\bibnamefont
  {Chan}}, \bibinfo {author} {\bibfnamefont {T.~M.}\ \bibnamefont {Alegre}},
  \bibinfo {author} {\bibfnamefont {A.~H.}\ \bibnamefont {Safavi-Naeini}},
  \bibinfo {author} {\bibfnamefont {J.~T.}\ \bibnamefont {Hill}}, \bibinfo
  {author} {\bibfnamefont {A.}~\bibnamefont {Krause}}, \bibinfo {author}
  {\bibfnamefont {S.}~\bibnamefont {Gr{\"o}blacher}}, \bibinfo {author}
  {\bibfnamefont {M.}~\bibnamefont {Aspelmeyer}}, \ and\ \bibinfo {author}
  {\bibfnamefont {O.}~\bibnamefont {Painter}},\ }\href@noop {} {\bibfield
  {journal} {\bibinfo  {journal} {Nature}\ }\textbf {\bibinfo {volume} {478}},\
  \bibinfo {pages} {89} (\bibinfo {year} {2011})}\BibitemShut {NoStop}%
\bibitem [{\citenamefont {Teufel}\ \emph {et~al.}(2011)\citenamefont {Teufel},
  \citenamefont {Donner}, \citenamefont {Li}, \citenamefont {Harlow},
  \citenamefont {Allman}, \citenamefont {Cicak}, \citenamefont {Sirois},
  \citenamefont {Whittaker}, \citenamefont {Lehnert},\ and\ \citenamefont
  {Simmonds}}]{teufel2011sideband}%
  \BibitemOpen
  \bibfield  {author} {\bibinfo {author} {\bibfnamefont {J.~D.}\ \bibnamefont
  {Teufel}}, \bibinfo {author} {\bibfnamefont {T.}~\bibnamefont {Donner}},
  \bibinfo {author} {\bibfnamefont {D.}~\bibnamefont {Li}}, \bibinfo {author}
  {\bibfnamefont {J.~W.}\ \bibnamefont {Harlow}}, \bibinfo {author}
  {\bibfnamefont {M.~S.}\ \bibnamefont {Allman}}, \bibinfo {author}
  {\bibfnamefont {K.}~\bibnamefont {Cicak}}, \bibinfo {author} {\bibfnamefont
  {A.~J.}\ \bibnamefont {Sirois}}, \bibinfo {author} {\bibfnamefont {J.~D.}\
  \bibnamefont {Whittaker}}, \bibinfo {author} {\bibfnamefont {K.~W.}\
  \bibnamefont {Lehnert}}, \ and\ \bibinfo {author} {\bibfnamefont {R.~W.}\
  \bibnamefont {Simmonds}},\ }\href {\doibase
  https://doi.org/10.1038/nature10261} {\bibfield  {journal} {\bibinfo
  {journal} {Nature}\ }\textbf {\bibinfo {volume} {475}},\ \bibinfo {pages}
  {359} (\bibinfo {year} {2011})}\BibitemShut {NoStop}%
\bibitem [{\citenamefont {Brooks}\ \emph {et~al.}(2012)\citenamefont {Brooks},
  \citenamefont {Botter}, \citenamefont {Schreppler}, \citenamefont {Purdy},
  \citenamefont {Brahms},\ and\ \citenamefont {Stamper-Kurn}}]{brooks2012non}%
  \BibitemOpen
  \bibfield  {author} {\bibinfo {author} {\bibfnamefont {D.~W.}\ \bibnamefont
  {Brooks}}, \bibinfo {author} {\bibfnamefont {T.}~\bibnamefont {Botter}},
  \bibinfo {author} {\bibfnamefont {S.}~\bibnamefont {Schreppler}}, \bibinfo
  {author} {\bibfnamefont {T.~P.}\ \bibnamefont {Purdy}}, \bibinfo {author}
  {\bibfnamefont {N.}~\bibnamefont {Brahms}}, \ and\ \bibinfo {author}
  {\bibfnamefont {D.~M.}\ \bibnamefont {Stamper-Kurn}},\ }\href@noop {}
  {\bibfield  {journal} {\bibinfo  {journal} {Nature}\ }\textbf {\bibinfo
  {volume} {488}},\ \bibinfo {pages} {476} (\bibinfo {year}
  {2012})}\BibitemShut {NoStop}%
\bibitem [{\citenamefont {Safavi-Naeini}\ \emph {et~al.}(2013)\citenamefont
  {Safavi-Naeini}, \citenamefont {Gr{\"o}blacher}, \citenamefont {Hill},
  \citenamefont {Chan}, \citenamefont {Aspelmeyer},\ and\ \citenamefont
  {Painter}}]{safavi2013squeezed}%
  \BibitemOpen
  \bibfield  {author} {\bibinfo {author} {\bibfnamefont {A.~H.}\ \bibnamefont
  {Safavi-Naeini}}, \bibinfo {author} {\bibfnamefont {S.}~\bibnamefont
  {Gr{\"o}blacher}}, \bibinfo {author} {\bibfnamefont {J.~T.}\ \bibnamefont
  {Hill}}, \bibinfo {author} {\bibfnamefont {J.}~\bibnamefont {Chan}}, \bibinfo
  {author} {\bibfnamefont {M.}~\bibnamefont {Aspelmeyer}}, \ and\ \bibinfo
  {author} {\bibfnamefont {O.}~\bibnamefont {Painter}},\ }\href@noop {}
  {\bibfield  {journal} {\bibinfo  {journal} {Nature}\ }\textbf {\bibinfo
  {volume} {500}},\ \bibinfo {pages} {185} (\bibinfo {year}
  {2013})}\BibitemShut {NoStop}%
\bibitem [{\citenamefont {Purdy}\ \emph {et~al.}(2013)\citenamefont {Purdy},
  \citenamefont {Yu}, \citenamefont {Peterson}, \citenamefont {Kampel},\ and\
  \citenamefont {Regal}}]{purdy2013strong}%
  \BibitemOpen
  \bibfield  {author} {\bibinfo {author} {\bibfnamefont {T.~P.}\ \bibnamefont
  {Purdy}}, \bibinfo {author} {\bibfnamefont {P.-L.}\ \bibnamefont {Yu}},
  \bibinfo {author} {\bibfnamefont {R.}~\bibnamefont {Peterson}}, \bibinfo
  {author} {\bibfnamefont {N.}~\bibnamefont {Kampel}}, \ and\ \bibinfo {author}
  {\bibfnamefont {C.}~\bibnamefont {Regal}},\ }\href@noop {} {\bibfield
  {journal} {\bibinfo  {journal} {Physical Review X}\ }\textbf {\bibinfo
  {volume} {3}},\ \bibinfo {pages} {031012} (\bibinfo {year}
  {2013})}\BibitemShut {NoStop}%
\bibitem [{\citenamefont {Palomaki}\ \emph {et~al.}(2013)\citenamefont
  {Palomaki}, \citenamefont {Teufel}, \citenamefont {Simmonds},\ and\
  \citenamefont {Lehnert}}]{palomaki2013entangling}%
  \BibitemOpen
  \bibfield  {author} {\bibinfo {author} {\bibfnamefont {T.~A.}\ \bibnamefont
  {Palomaki}}, \bibinfo {author} {\bibfnamefont {J.~D.}\ \bibnamefont
  {Teufel}}, \bibinfo {author} {\bibfnamefont {R.~W.}\ \bibnamefont
  {Simmonds}}, \ and\ \bibinfo {author} {\bibfnamefont {K.~W.}\ \bibnamefont
  {Lehnert}},\ }\href {\doibase 10.1126/science.1244563} {\bibfield  {journal}
  {\bibinfo  {journal} {Science}\ }\textbf {\bibinfo {volume} {342}},\ \bibinfo
  {pages} {710} (\bibinfo {year} {2013})}\BibitemShut {NoStop}%
\bibitem [{\citenamefont {Reed}\ \emph {et~al.}(2017)\citenamefont {Reed},
  \citenamefont {Mayer}, \citenamefont {Teufel}, \citenamefont {Burkhart},
  \citenamefont {Pfaff}, \citenamefont {Reagor}, \citenamefont {Sletten},
  \citenamefont {Ma}, \citenamefont {Schoelkopf}, \citenamefont {Knill} \emph
  {et~al.}}]{reed2017faithful}%
  \BibitemOpen
  \bibfield  {author} {\bibinfo {author} {\bibfnamefont {A.}~\bibnamefont
  {Reed}}, \bibinfo {author} {\bibfnamefont {K.}~\bibnamefont {Mayer}},
  \bibinfo {author} {\bibfnamefont {J.}~\bibnamefont {Teufel}}, \bibinfo
  {author} {\bibfnamefont {L.}~\bibnamefont {Burkhart}}, \bibinfo {author}
  {\bibfnamefont {W.}~\bibnamefont {Pfaff}}, \bibinfo {author} {\bibfnamefont
  {M.}~\bibnamefont {Reagor}}, \bibinfo {author} {\bibfnamefont
  {L.}~\bibnamefont {Sletten}}, \bibinfo {author} {\bibfnamefont
  {X.}~\bibnamefont {Ma}}, \bibinfo {author} {\bibfnamefont {R.}~\bibnamefont
  {Schoelkopf}}, \bibinfo {author} {\bibfnamefont {E.}~\bibnamefont {Knill}},
  \emph {et~al.},\ }\href@noop {} {\bibfield  {journal} {\bibinfo  {journal}
  {Nature Physics}\ }\textbf {\bibinfo {volume} {13}},\ \bibinfo {pages} {1163}
  (\bibinfo {year} {2017})}\BibitemShut {NoStop}%
\bibitem [{\citenamefont {Hong}\ \emph {et~al.}(2017)\citenamefont {Hong},
  \citenamefont {Riedinger}, \citenamefont {Marinkovi{\'c}}, \citenamefont
  {Wallucks}, \citenamefont {Hofer}, \citenamefont {Norte}, \citenamefont
  {Aspelmeyer},\ and\ \citenamefont {Gr{\"o}blacher}}]{Hong2017handbury}%
  \BibitemOpen
  \bibfield  {author} {\bibinfo {author} {\bibfnamefont {S.}~\bibnamefont
  {Hong}}, \bibinfo {author} {\bibfnamefont {R.}~\bibnamefont {Riedinger}},
  \bibinfo {author} {\bibfnamefont {I.}~\bibnamefont {Marinkovi{\'c}}},
  \bibinfo {author} {\bibfnamefont {A.}~\bibnamefont {Wallucks}}, \bibinfo
  {author} {\bibfnamefont {S.~G.}\ \bibnamefont {Hofer}}, \bibinfo {author}
  {\bibfnamefont {R.~A.}\ \bibnamefont {Norte}}, \bibinfo {author}
  {\bibfnamefont {M.}~\bibnamefont {Aspelmeyer}}, \ and\ \bibinfo {author}
  {\bibfnamefont {S.}~\bibnamefont {Gr{\"o}blacher}},\ }\href {\doibase
  10.1126/science.aan7939} {\bibfield  {journal} {\bibinfo  {journal}
  {Science}\ }\textbf {\bibinfo {volume} {358}},\ \bibinfo {pages} {203}
  (\bibinfo {year} {2017})}\BibitemShut {NoStop}%
\bibitem [{\citenamefont {Meenehan}\ \emph {et~al.}(2014)\citenamefont
  {Meenehan}, \citenamefont {Cohen}, \citenamefont {Gr{\"o}blacher},
  \citenamefont {Hill}, \citenamefont {Safavi-Naeini}, \citenamefont
  {Aspelmeyer},\ and\ \citenamefont {Painter}}]{meenehan2014silicon}%
  \BibitemOpen
  \bibfield  {author} {\bibinfo {author} {\bibfnamefont {S.~M.}\ \bibnamefont
  {Meenehan}}, \bibinfo {author} {\bibfnamefont {J.~D.}\ \bibnamefont {Cohen}},
  \bibinfo {author} {\bibfnamefont {S.}~\bibnamefont {Gr{\"o}blacher}},
  \bibinfo {author} {\bibfnamefont {J.~T.}\ \bibnamefont {Hill}}, \bibinfo
  {author} {\bibfnamefont {A.~H.}\ \bibnamefont {Safavi-Naeini}}, \bibinfo
  {author} {\bibfnamefont {M.}~\bibnamefont {Aspelmeyer}}, \ and\ \bibinfo
  {author} {\bibfnamefont {O.}~\bibnamefont {Painter}},\ }\href@noop {}
  {\bibfield  {journal} {\bibinfo  {journal} {Physical Review A}\ }\textbf
  {\bibinfo {volume} {90}},\ \bibinfo {pages} {011803} (\bibinfo {year}
  {2014})}\BibitemShut {NoStop}%
\bibitem [{\citenamefont {Yuan}\ \emph {et~al.}(2015)\citenamefont {Yuan},
  \citenamefont {Singh}, \citenamefont {Blanter},\ and\ \citenamefont
  {Steele}}]{yuan2015large}%
  \BibitemOpen
  \bibfield  {author} {\bibinfo {author} {\bibfnamefont {M.}~\bibnamefont
  {Yuan}}, \bibinfo {author} {\bibfnamefont {V.}~\bibnamefont {Singh}},
  \bibinfo {author} {\bibfnamefont {Y.~M.}\ \bibnamefont {Blanter}}, \ and\
  \bibinfo {author} {\bibfnamefont {G.~A.}\ \bibnamefont {Steele}},\
  }\href@noop {} {\bibfield  {journal} {\bibinfo  {journal} {Nature
  communications}\ }\textbf {\bibinfo {volume} {6}},\ \bibinfo {pages} {8491}
  (\bibinfo {year} {2015})}\BibitemShut {NoStop}%
\bibitem [{\citenamefont {Rabl}(2011)}]{rabl2011photon}%
  \BibitemOpen
  \bibfield  {author} {\bibinfo {author} {\bibfnamefont {P.}~\bibnamefont
  {Rabl}},\ }\href {\doibase 10.1103/PhysRevLett.107.063601} {\bibfield
  {journal} {\bibinfo  {journal} {Phys. Rev. Lett.}\ }\textbf {\bibinfo
  {volume} {107}},\ \bibinfo {pages} {063601} (\bibinfo {year}
  {2011})}\BibitemShut {NoStop}%
\bibitem [{\citenamefont {Nunnenkamp}\ \emph {et~al.}(2011)\citenamefont
  {Nunnenkamp}, \citenamefont {B\o{}rkje},\ and\ \citenamefont
  {Girvin}}]{nunnenkamp2011single}%
  \BibitemOpen
  \bibfield  {author} {\bibinfo {author} {\bibfnamefont {A.}~\bibnamefont
  {Nunnenkamp}}, \bibinfo {author} {\bibfnamefont {K.}~\bibnamefont
  {B\o{}rkje}}, \ and\ \bibinfo {author} {\bibfnamefont {S.~M.}\ \bibnamefont
  {Girvin}},\ }\href {\doibase 10.1103/PhysRevLett.107.063602} {\bibfield
  {journal} {\bibinfo  {journal} {Phys. Rev. Lett.}\ }\textbf {\bibinfo
  {volume} {107}},\ \bibinfo {pages} {063602} (\bibinfo {year}
  {2011})}\BibitemShut {NoStop}%
\bibitem [{\citenamefont {Qian}\ \emph {et~al.}(2012)\citenamefont {Qian},
  \citenamefont {Clerk}, \citenamefont {Hammerer},\ and\ \citenamefont
  {Marquardt}}]{qian2012quantum}%
  \BibitemOpen
  \bibfield  {author} {\bibinfo {author} {\bibfnamefont {J.}~\bibnamefont
  {Qian}}, \bibinfo {author} {\bibfnamefont {A.~A.}\ \bibnamefont {Clerk}},
  \bibinfo {author} {\bibfnamefont {K.}~\bibnamefont {Hammerer}}, \ and\
  \bibinfo {author} {\bibfnamefont {F.}~\bibnamefont {Marquardt}},\ }\href
  {\doibase 10.1103/PhysRevLett.109.253601} {\bibfield  {journal} {\bibinfo
  {journal} {Phys. Rev. Lett.}\ }\textbf {\bibinfo {volume} {109}},\ \bibinfo
  {pages} {253601} (\bibinfo {year} {2012})}\BibitemShut {NoStop}%
\bibitem [{\citenamefont {Garziano}\ \emph {et~al.}(2015)\citenamefont
  {Garziano}, \citenamefont {Stassi}, \citenamefont {Macr\'{\i}}, \citenamefont
  {Savasta},\ and\ \citenamefont {Di~Stefano}}]{garziano2015single}%
  \BibitemOpen
  \bibfield  {author} {\bibinfo {author} {\bibfnamefont {L.}~\bibnamefont
  {Garziano}}, \bibinfo {author} {\bibfnamefont {R.}~\bibnamefont {Stassi}},
  \bibinfo {author} {\bibfnamefont {V.}~\bibnamefont {Macr\'{\i}}}, \bibinfo
  {author} {\bibfnamefont {S.}~\bibnamefont {Savasta}}, \ and\ \bibinfo
  {author} {\bibfnamefont {O.}~\bibnamefont {Di~Stefano}},\ }\href {\doibase
  10.1103/PhysRevA.91.023809} {\bibfield  {journal} {\bibinfo  {journal} {Phys.
  Rev. A}\ }\textbf {\bibinfo {volume} {91}},\ \bibinfo {pages} {023809}
  (\bibinfo {year} {2015})}\BibitemShut {NoStop}%
\bibitem [{\citenamefont {Liao}\ and\ \citenamefont
  {Tian}(2016)}]{liao2016macroscopic}%
  \BibitemOpen
  \bibfield  {author} {\bibinfo {author} {\bibfnamefont {J.-Q.}\ \bibnamefont
  {Liao}}\ and\ \bibinfo {author} {\bibfnamefont {L.}~\bibnamefont {Tian}},\
  }\href {\doibase 10.1103/PhysRevLett.116.163602} {\bibfield  {journal}
  {\bibinfo  {journal} {Phys. Rev. Lett.}\ }\textbf {\bibinfo {volume} {116}},\
  \bibinfo {pages} {163602} (\bibinfo {year} {2016})}\BibitemShut {NoStop}%
\bibitem [{\citenamefont {Zhou}\ and\ \citenamefont
  {Mizel}(2006)}]{zhou2006nonlinear}%
  \BibitemOpen
  \bibfield  {author} {\bibinfo {author} {\bibfnamefont {X.}~\bibnamefont
  {Zhou}}\ and\ \bibinfo {author} {\bibfnamefont {A.}~\bibnamefont {Mizel}},\
  }\href {\doibase 10.1103/PhysRevLett.97.267201} {\bibfield  {journal}
  {\bibinfo  {journal} {Phys. Rev. Lett.}\ }\textbf {\bibinfo {volume} {97}},\
  \bibinfo {pages} {267201} (\bibinfo {year} {2006})}\BibitemShut {NoStop}%
\bibitem [{\citenamefont {Buks}\ and\ \citenamefont
  {Blencowe}(2006)}]{buks2006decoherence}%
  \BibitemOpen
  \bibfield  {author} {\bibinfo {author} {\bibfnamefont {E.}~\bibnamefont
  {Buks}}\ and\ \bibinfo {author} {\bibfnamefont {M.~P.}\ \bibnamefont
  {Blencowe}},\ }\href {\doibase 10.1103/PhysRevB.74.174504} {\bibfield
  {journal} {\bibinfo  {journal} {Phys. Rev. B}\ }\textbf {\bibinfo {volume}
  {74}},\ \bibinfo {pages} {174504} (\bibinfo {year} {2006})}\BibitemShut
  {NoStop}%
\bibitem [{\citenamefont {Blencowe}\ and\ \citenamefont
  {Buks}(2007)}]{blencowe2007quantum}%
  \BibitemOpen
  \bibfield  {author} {\bibinfo {author} {\bibfnamefont {M.~P.}\ \bibnamefont
  {Blencowe}}\ and\ \bibinfo {author} {\bibfnamefont {E.}~\bibnamefont
  {Buks}},\ }\href {\doibase 10.1103/PhysRevB.76.014511} {\bibfield  {journal}
  {\bibinfo  {journal} {Phys. Rev. B}\ }\textbf {\bibinfo {volume} {76}},\
  \bibinfo {pages} {014511} (\bibinfo {year} {2007})}\BibitemShut {NoStop}%
\bibitem [{\citenamefont {Nation}\ \emph {et~al.}(2008)\citenamefont {Nation},
  \citenamefont {Blencowe},\ and\ \citenamefont {Buks}}]{nation2008quantum}%
  \BibitemOpen
  \bibfield  {author} {\bibinfo {author} {\bibfnamefont {P.~D.}\ \bibnamefont
  {Nation}}, \bibinfo {author} {\bibfnamefont {M.~P.}\ \bibnamefont
  {Blencowe}}, \ and\ \bibinfo {author} {\bibfnamefont {E.}~\bibnamefont
  {Buks}},\ }\href {\doibase 10.1103/PhysRevB.78.104516} {\bibfield  {journal}
  {\bibinfo  {journal} {Phys. Rev. B}\ }\textbf {\bibinfo {volume} {78}},\
  \bibinfo {pages} {104516} (\bibinfo {year} {2008})}\BibitemShut {NoStop}%
\bibitem [{\citenamefont {Etaki}\ \emph {et~al.}(2008)\citenamefont {Etaki},
  \citenamefont {Poot}, \citenamefont {Mahboob}, \citenamefont {Onomitsu},
  \citenamefont {Yamaguchi},\ and\ \citenamefont {Van~der
  Zant}}]{etaki2008motion}%
  \BibitemOpen
  \bibfield  {author} {\bibinfo {author} {\bibfnamefont {S.}~\bibnamefont
  {Etaki}}, \bibinfo {author} {\bibfnamefont {M.}~\bibnamefont {Poot}},
  \bibinfo {author} {\bibfnamefont {I.}~\bibnamefont {Mahboob}}, \bibinfo
  {author} {\bibfnamefont {K.}~\bibnamefont {Onomitsu}}, \bibinfo {author}
  {\bibfnamefont {H.}~\bibnamefont {Yamaguchi}}, \ and\ \bibinfo {author}
  {\bibfnamefont {H.}~\bibnamefont {Van~der Zant}},\ }\href@noop {} {\bibfield
  {journal} {\bibinfo  {journal} {Nature Physics}\ }\textbf {\bibinfo {volume}
  {4}},\ \bibinfo {pages} {785} (\bibinfo {year} {2008})}\BibitemShut {NoStop}%
\bibitem [{\citenamefont {Nation}(2013)}]{nation2013nonclassical}%
  \BibitemOpen
  \bibfield  {author} {\bibinfo {author} {\bibfnamefont {P.~D.}\ \bibnamefont
  {Nation}},\ }\href {\doibase 10.1103/PhysRevA.88.053828} {\bibfield
  {journal} {\bibinfo  {journal} {Phys. Rev. A}\ }\textbf {\bibinfo {volume}
  {88}},\ \bibinfo {pages} {053828} (\bibinfo {year} {2013})}\BibitemShut
  {NoStop}%
\bibitem [{\citenamefont {Rimberg}\ \emph {et~al.}(2014)\citenamefont
  {Rimberg}, \citenamefont {Blencowe}, \citenamefont {Armour},\ and\
  \citenamefont {Nation}}]{rimberg2014cavity}%
  \BibitemOpen
  \bibfield  {author} {\bibinfo {author} {\bibfnamefont {A.~J.}\ \bibnamefont
  {Rimberg}}, \bibinfo {author} {\bibfnamefont {M.~P.}\ \bibnamefont
  {Blencowe}}, \bibinfo {author} {\bibfnamefont {A.~D.}\ \bibnamefont
  {Armour}}, \ and\ \bibinfo {author} {\bibfnamefont {P.~D.}\ \bibnamefont
  {Nation}},\ }\href {\doibase 10.1088/1367-2630/16/5/055008} {\bibfield
  {journal} {\bibinfo  {journal} {New Journal of Physics}\ }\textbf {\bibinfo
  {volume} {16}},\ \bibinfo {pages} {055008} (\bibinfo {year}
  {2014})}\BibitemShut {NoStop}%
\bibitem [{\citenamefont {Nation}\ \emph {et~al.}(2016)\citenamefont {Nation},
  \citenamefont {Suh},\ and\ \citenamefont {Blencowe}}]{nation2016ultrastrong}%
  \BibitemOpen
  \bibfield  {author} {\bibinfo {author} {\bibfnamefont {P.~D.}\ \bibnamefont
  {Nation}}, \bibinfo {author} {\bibfnamefont {J.}~\bibnamefont {Suh}}, \ and\
  \bibinfo {author} {\bibfnamefont {M.~P.}\ \bibnamefont {Blencowe}},\ }\href
  {\doibase 10.1103/PhysRevA.93.022510} {\bibfield  {journal} {\bibinfo
  {journal} {Phys. Rev. A}\ }\textbf {\bibinfo {volume} {93}},\ \bibinfo
  {pages} {022510} (\bibinfo {year} {2016})}\BibitemShut {NoStop}%
\bibitem [{\citenamefont {Shevchuk}\ \emph {et~al.}(2017)\citenamefont
  {Shevchuk}, \citenamefont {Steele},\ and\ \citenamefont
  {Blanter}}]{shevchuk2017strong}%
  \BibitemOpen
  \bibfield  {author} {\bibinfo {author} {\bibfnamefont {O.}~\bibnamefont
  {Shevchuk}}, \bibinfo {author} {\bibfnamefont {G.~A.}\ \bibnamefont
  {Steele}}, \ and\ \bibinfo {author} {\bibfnamefont {Y.~M.}\ \bibnamefont
  {Blanter}},\ }\href {\doibase 10.1103/PhysRevB.96.014508} {\bibfield
  {journal} {\bibinfo  {journal} {Phys. Rev. B}\ }\textbf {\bibinfo {volume}
  {96}},\ \bibinfo {pages} {014508} (\bibinfo {year} {2017})}\BibitemShut
  {NoStop}%
\bibitem [{\citenamefont {Khosla}\ \emph {et~al.}(2018)\citenamefont {Khosla},
  \citenamefont {Vanner}, \citenamefont {Ares},\ and\ \citenamefont
  {Laird}}]{khosla2018displacemon}%
  \BibitemOpen
  \bibfield  {author} {\bibinfo {author} {\bibfnamefont {K.~E.}\ \bibnamefont
  {Khosla}}, \bibinfo {author} {\bibfnamefont {M.~R.}\ \bibnamefont {Vanner}},
  \bibinfo {author} {\bibfnamefont {N.}~\bibnamefont {Ares}}, \ and\ \bibinfo
  {author} {\bibfnamefont {E.~A.}\ \bibnamefont {Laird}},\ }\href {\doibase
  10.1103/PhysRevX.8.021052} {\bibfield  {journal} {\bibinfo  {journal} {Phys.
  Rev. X}\ }\textbf {\bibinfo {volume} {8}},\ \bibinfo {pages} {021052}
  (\bibinfo {year} {2018})}\BibitemShut {NoStop}%
\bibitem [{\citenamefont {Rodrigues}\ \emph {et~al.}(2019)\citenamefont
  {Rodrigues}, \citenamefont {Bothner},\ and\ \citenamefont
  {Steele}}]{rodrigues2019coupling}%
  \BibitemOpen
  \bibfield  {author} {\bibinfo {author} {\bibfnamefont {I.}~\bibnamefont
  {Rodrigues}}, \bibinfo {author} {\bibfnamefont {D.}~\bibnamefont {Bothner}},
  \ and\ \bibinfo {author} {\bibfnamefont {G.}~\bibnamefont {Steele}},\
  }\href@noop {} {\bibfield  {journal} {\bibinfo  {journal} {Nature
  communications}\ }\textbf {\bibinfo {volume} {10}},\ \bibinfo {pages} {1}
  (\bibinfo {year} {2019})}\BibitemShut {NoStop}%
\bibitem [{\citenamefont {Koch}\ \emph {et~al.}(2007)\citenamefont {Koch},
  \citenamefont {Yu}, \citenamefont {Gambetta}, \citenamefont {Houck},
  \citenamefont {Schuster}, \citenamefont {Majer}, \citenamefont {Blais},
  \citenamefont {Devoret}, \citenamefont {Girvin},\ and\ \citenamefont
  {Schoelkopf}}]{koch2007charge}%
  \BibitemOpen
  \bibfield  {author} {\bibinfo {author} {\bibfnamefont {J.}~\bibnamefont
  {Koch}}, \bibinfo {author} {\bibfnamefont {T.~M.}\ \bibnamefont {Yu}},
  \bibinfo {author} {\bibfnamefont {J.}~\bibnamefont {Gambetta}}, \bibinfo
  {author} {\bibfnamefont {A.~A.}\ \bibnamefont {Houck}}, \bibinfo {author}
  {\bibfnamefont {D.~I.}\ \bibnamefont {Schuster}}, \bibinfo {author}
  {\bibfnamefont {J.}~\bibnamefont {Majer}}, \bibinfo {author} {\bibfnamefont
  {A.}~\bibnamefont {Blais}}, \bibinfo {author} {\bibfnamefont {M.~H.}\
  \bibnamefont {Devoret}}, \bibinfo {author} {\bibfnamefont {S.~M.}\
  \bibnamefont {Girvin}}, \ and\ \bibinfo {author} {\bibfnamefont {R.~J.}\
  \bibnamefont {Schoelkopf}},\ }\href {\doibase 10.1103/PhysRevA.76.042319}
  {\bibfield  {journal} {\bibinfo  {journal} {Phys. Rev. A}\ }\textbf {\bibinfo
  {volume} {76}},\ \bibinfo {pages} {042319} (\bibinfo {year}
  {2007})}\BibitemShut {NoStop}%
\bibitem [{\citenamefont {Kounalakis}\ \emph {et~al.}(2018)\citenamefont
  {Kounalakis}, \citenamefont {Dickel}, \citenamefont {Bruno}, \citenamefont
  {Langford},\ and\ \citenamefont {Steele}}]{kounalakis2018tuneable}%
  \BibitemOpen
  \bibfield  {author} {\bibinfo {author} {\bibfnamefont {M.}~\bibnamefont
  {Kounalakis}}, \bibinfo {author} {\bibfnamefont {C.}~\bibnamefont {Dickel}},
  \bibinfo {author} {\bibfnamefont {A.}~\bibnamefont {Bruno}}, \bibinfo
  {author} {\bibfnamefont {N.}~\bibnamefont {Langford}}, \ and\ \bibinfo
  {author} {\bibfnamefont {G.}~\bibnamefont {Steele}},\ }\href
  {https://doi.org/10.1038/s41534-018-0088-9} {\bibfield  {journal} {\bibinfo
  {journal} {npj Quantum Information}\ }\textbf {\bibinfo {volume} {4}},\
  \bibinfo {pages} {38} (\bibinfo {year} {2018})}\BibitemShut {NoStop}%
\bibitem [{\citenamefont {Clarke}\ and\ \citenamefont
  {Braginski}(2004)}]{clarke2004squid}%
  \BibitemOpen
  \bibfield  {author} {\bibinfo {author} {\bibfnamefont {J.}~\bibnamefont
  {Clarke}}\ and\ \bibinfo {author} {\bibfnamefont {A.~I.}\ \bibnamefont
  {Braginski}},\ }\href@noop {} {\emph {\bibinfo {title} {The SQUID
  handbook}}}\ (\bibinfo  {publisher} {Wiley Online Library},\ \bibinfo {year}
  {2004})\BibitemShut {NoStop}%
\bibitem [{\citenamefont {Johansson}\ \emph {et~al.}(2012)\citenamefont
  {Johansson}, \citenamefont {Nation},\ and\ \citenamefont
  {Nori}}]{johansson2012qutip}%
  \BibitemOpen
  \bibfield  {author} {\bibinfo {author} {\bibfnamefont {J.}~\bibnamefont
  {Johansson}}, \bibinfo {author} {\bibfnamefont {P.}~\bibnamefont {Nation}}, \
  and\ \bibinfo {author} {\bibfnamefont {F.}~\bibnamefont {Nori}},\ }\href@noop
  {} {\bibfield  {journal} {\bibinfo  {journal} {Computer Physics
  Communications}\ }\textbf {\bibinfo {volume} {183}},\ \bibinfo {pages} {1760}
  (\bibinfo {year} {2012})}\BibitemShut {NoStop}%
\bibitem [{\citenamefont {Schneider}\ \emph {et~al.}(2019)\citenamefont
  {Schneider}, \citenamefont {Wolz}, \citenamefont {Pfirrmann}, \citenamefont
  {Spiecker}, \citenamefont {Rotzinger}, \citenamefont {Ustinov},\ and\
  \citenamefont {Weides}}]{schneider2019transmon}%
  \BibitemOpen
  \bibfield  {author} {\bibinfo {author} {\bibfnamefont {A.}~\bibnamefont
  {Schneider}}, \bibinfo {author} {\bibfnamefont {T.}~\bibnamefont {Wolz}},
  \bibinfo {author} {\bibfnamefont {M.}~\bibnamefont {Pfirrmann}}, \bibinfo
  {author} {\bibfnamefont {M.}~\bibnamefont {Spiecker}}, \bibinfo {author}
  {\bibfnamefont {H.}~\bibnamefont {Rotzinger}}, \bibinfo {author}
  {\bibfnamefont {A.~V.}\ \bibnamefont {Ustinov}}, \ and\ \bibinfo {author}
  {\bibfnamefont {M.}~\bibnamefont {Weides}},\ }\href {\doibase
  10.1103/PhysRevResearch.1.023003} {\bibfield  {journal} {\bibinfo  {journal}
  {Phys. Rev. Research}\ }\textbf {\bibinfo {volume} {1}},\ \bibinfo {pages}
  {023003} (\bibinfo {year} {2019})}\BibitemShut {NoStop}%
\bibitem [{\citenamefont {Rist{\`e}}\ \emph {et~al.}(2013)\citenamefont
  {Rist{\`e}}, \citenamefont {Bultink}, \citenamefont {Tiggelman},
  \citenamefont {Schouten}, \citenamefont {Lehnert},\ and\ \citenamefont
  {DiCarlo}}]{riste2013millisecond}%
  \BibitemOpen
  \bibfield  {author} {\bibinfo {author} {\bibfnamefont {D.}~\bibnamefont
  {Rist{\`e}}}, \bibinfo {author} {\bibfnamefont {C.~C.}\ \bibnamefont
  {Bultink}}, \bibinfo {author} {\bibfnamefont {M.~J.}\ \bibnamefont
  {Tiggelman}}, \bibinfo {author} {\bibfnamefont {R.~N.}\ \bibnamefont
  {Schouten}}, \bibinfo {author} {\bibfnamefont {K.~W.}\ \bibnamefont
  {Lehnert}}, \ and\ \bibinfo {author} {\bibfnamefont {L.}~\bibnamefont
  {DiCarlo}},\ }\href {http://dx.doi.org/10.1038/ncomms2936} {\bibfield
  {journal} {\bibinfo  {journal} {Nature Communications}\ }\textbf {\bibinfo
  {volume} {4}},\ \bibinfo {pages} {1913} (\bibinfo {year} {2013})}\BibitemShut
  {NoStop}%
\bibitem [{\citenamefont {Gely}\ \emph {et~al.}(2019)\citenamefont {Gely},
  \citenamefont {Kounalakis}, \citenamefont {Dickel}, \citenamefont {Dalle},
  \citenamefont {Vatr{\'e}}, \citenamefont {Baker}, \citenamefont {Jenkins},\
  and\ \citenamefont {Steele}}]{gely2019observation}%
  \BibitemOpen
  \bibfield  {author} {\bibinfo {author} {\bibfnamefont {M.~F.}\ \bibnamefont
  {Gely}}, \bibinfo {author} {\bibfnamefont {M.}~\bibnamefont {Kounalakis}},
  \bibinfo {author} {\bibfnamefont {C.}~\bibnamefont {Dickel}}, \bibinfo
  {author} {\bibfnamefont {J.}~\bibnamefont {Dalle}}, \bibinfo {author}
  {\bibfnamefont {R.}~\bibnamefont {Vatr{\'e}}}, \bibinfo {author}
  {\bibfnamefont {B.}~\bibnamefont {Baker}}, \bibinfo {author} {\bibfnamefont
  {M.~D.}\ \bibnamefont {Jenkins}}, \ and\ \bibinfo {author} {\bibfnamefont
  {G.~A.}\ \bibnamefont {Steele}},\ }\href {\doibase 10.1126/science.aaw3101}
  {\bibfield  {journal} {\bibinfo  {journal} {Science}\ }\textbf {\bibinfo
  {volume} {363}},\ \bibinfo {pages} {1072} (\bibinfo {year}
  {2019})}\BibitemShut {NoStop}%
\bibitem [{\citenamefont {Lescanne}\ \emph {et~al.}(2019)\citenamefont
  {Lescanne}, \citenamefont {Verney}, \citenamefont {Ficheux}, \citenamefont
  {Devoret}, \citenamefont {Huard}, \citenamefont {Mirrahimi},\ and\
  \citenamefont {Leghtas}}]{lescanne2019observing}%
  \BibitemOpen
  \bibfield  {author} {\bibinfo {author} {\bibfnamefont {R.}~\bibnamefont
  {Lescanne}}, \bibinfo {author} {\bibfnamefont {L.}~\bibnamefont {Verney}},
  \bibinfo {author} {\bibfnamefont {Q.}~\bibnamefont {Ficheux}}, \bibinfo
  {author} {\bibfnamefont {M.~H.}\ \bibnamefont {Devoret}}, \bibinfo {author}
  {\bibfnamefont {B.}~\bibnamefont {Huard}}, \bibinfo {author} {\bibfnamefont
  {M.}~\bibnamefont {Mirrahimi}}, \ and\ \bibinfo {author} {\bibfnamefont
  {Z.}~\bibnamefont {Leghtas}},\ }\href {\doibase
  10.1103/PhysRevApplied.11.014030} {\bibfield  {journal} {\bibinfo  {journal}
  {Phys. Rev. Applied}\ }\textbf {\bibinfo {volume} {11}},\ \bibinfo {pages}
  {014030} (\bibinfo {year} {2019})}\BibitemShut {NoStop}%
\bibitem [{\citenamefont {Kounalakis}\ \emph {et~al.}(2019)\citenamefont
  {Kounalakis}, \citenamefont {Blanter},\ and\ \citenamefont
  {Steele}}]{kounalakis2019synthesizing}%
  \BibitemOpen
  \bibfield  {author} {\bibinfo {author} {\bibfnamefont {M.}~\bibnamefont
  {Kounalakis}}, \bibinfo {author} {\bibfnamefont {Y.~M.}\ \bibnamefont
  {Blanter}}, \ and\ \bibinfo {author} {\bibfnamefont {G.~A.}\ \bibnamefont
  {Steele}},\ }\href {\doibase 10.1038/s41534-019-0219-y} {\bibfield  {journal}
  {\bibinfo  {journal} {npj Quantum Information}\ }\textbf {\bibinfo {volume}
  {5}},\ \bibinfo {pages} {100} (\bibinfo {year} {2019})}\BibitemShut {NoStop}%
\bibitem [{\citenamefont {Hu}\ \emph {et~al.}(2015)\citenamefont {Hu},
  \citenamefont {Huang}, \citenamefont {Liao}, \citenamefont {Tian},\ and\
  \citenamefont {Goan}}]{hu2015quantum}%
  \BibitemOpen
  \bibfield  {author} {\bibinfo {author} {\bibfnamefont {D.}~\bibnamefont
  {Hu}}, \bibinfo {author} {\bibfnamefont {S.-Y.}\ \bibnamefont {Huang}},
  \bibinfo {author} {\bibfnamefont {J.-Q.}\ \bibnamefont {Liao}}, \bibinfo
  {author} {\bibfnamefont {L.}~\bibnamefont {Tian}}, \ and\ \bibinfo {author}
  {\bibfnamefont {H.-S.}\ \bibnamefont {Goan}},\ }\href {\doibase
  10.1103/PhysRevA.91.013812} {\bibfield  {journal} {\bibinfo  {journal} {Phys.
  Rev. A}\ }\textbf {\bibinfo {volume} {91}},\ \bibinfo {pages} {013812}
  (\bibinfo {year} {2015})}\BibitemShut {NoStop}%
\bibitem [{\citenamefont {Naseem}\ \emph {et~al.}(2018)\citenamefont {Naseem},
  \citenamefont {Xuereb},\ and\ \citenamefont {M\"ustecapl\ifmmode \imath \else
  \i \fi{}o\ifmmode~\breve{g}\else \u{g}\fi{}lu}}]{naseem2018thermodynamic}%
  \BibitemOpen
  \bibfield  {author} {\bibinfo {author} {\bibfnamefont {M.~T.}\ \bibnamefont
  {Naseem}}, \bibinfo {author} {\bibfnamefont {A.}~\bibnamefont {Xuereb}}, \
  and\ \bibinfo {author} {\bibfnamefont {O.~E.}\ \bibnamefont
  {M\"ustecapl\ifmmode \imath \else \i \fi{}o\ifmmode~\breve{g}\else
  \u{g}\fi{}lu}},\ }\href {\doibase 10.1103/PhysRevA.98.052123} {\bibfield
  {journal} {\bibinfo  {journal} {Phys. Rev. A}\ }\textbf {\bibinfo {volume}
  {98}},\ \bibinfo {pages} {052123} (\bibinfo {year} {2018})}\BibitemShut
  {NoStop}%
\bibitem [{\citenamefont {Beaudoin}\ \emph {et~al.}(2011)\citenamefont
  {Beaudoin}, \citenamefont {Gambetta},\ and\ \citenamefont
  {Blais}}]{beaudoin2011dissipation}%
  \BibitemOpen
  \bibfield  {author} {\bibinfo {author} {\bibfnamefont {F.}~\bibnamefont
  {Beaudoin}}, \bibinfo {author} {\bibfnamefont {J.~M.}\ \bibnamefont
  {Gambetta}}, \ and\ \bibinfo {author} {\bibfnamefont {A.}~\bibnamefont
  {Blais}},\ }\href {\doibase 10.1103/PhysRevA.84.043832} {\bibfield  {journal}
  {\bibinfo  {journal} {Phys. Rev. A}\ }\textbf {\bibinfo {volume} {84}},\
  \bibinfo {pages} {043832} (\bibinfo {year} {2011})}\BibitemShut {NoStop}%
\bibitem [{\citenamefont {Settineri}\ \emph {et~al.}(2018)\citenamefont
  {Settineri}, \citenamefont {Macr\'{\i}}, \citenamefont {Ridolfo},
  \citenamefont {Di~Stefano}, \citenamefont {Kockum}, \citenamefont {Nori},\
  and\ \citenamefont {Savasta}}]{settineri2018dissipation}%
  \BibitemOpen
  \bibfield  {author} {\bibinfo {author} {\bibfnamefont {A.}~\bibnamefont
  {Settineri}}, \bibinfo {author} {\bibfnamefont {V.}~\bibnamefont
  {Macr\'{\i}}}, \bibinfo {author} {\bibfnamefont {A.}~\bibnamefont {Ridolfo}},
  \bibinfo {author} {\bibfnamefont {O.}~\bibnamefont {Di~Stefano}}, \bibinfo
  {author} {\bibfnamefont {A.~F.}\ \bibnamefont {Kockum}}, \bibinfo {author}
  {\bibfnamefont {F.}~\bibnamefont {Nori}}, \ and\ \bibinfo {author}
  {\bibfnamefont {S.}~\bibnamefont {Savasta}},\ }\href {\doibase
  10.1103/PhysRevA.98.053834} {\bibfield  {journal} {\bibinfo  {journal} {Phys.
  Rev. A}\ }\textbf {\bibinfo {volume} {98}},\ \bibinfo {pages} {053834}
  (\bibinfo {year} {2018})}\BibitemShut {NoStop}%
\bibitem [{\citenamefont {Nunnenkamp}\ \emph {et~al.}(2012)\citenamefont
  {Nunnenkamp}, \citenamefont {B\o{}rkje},\ and\ \citenamefont
  {Girvin}}]{nunnenkamp2012cooling}%
  \BibitemOpen
  \bibfield  {author} {\bibinfo {author} {\bibfnamefont {A.}~\bibnamefont
  {Nunnenkamp}}, \bibinfo {author} {\bibfnamefont {K.}~\bibnamefont
  {B\o{}rkje}}, \ and\ \bibinfo {author} {\bibfnamefont {S.~M.}\ \bibnamefont
  {Girvin}},\ }\href {\doibase 10.1103/PhysRevA.85.051803} {\bibfield
  {journal} {\bibinfo  {journal} {Phys. Rev. A}\ }\textbf {\bibinfo {volume}
  {85}},\ \bibinfo {pages} {051803} (\bibinfo {year} {2012})}\BibitemShut
  {NoStop}%
\bibitem [{\citenamefont {Brune}\ \emph {et~al.}(1992)\citenamefont {Brune},
  \citenamefont {Haroche}, \citenamefont {Raimond}, \citenamefont
  {Davidovich},\ and\ \citenamefont {Zagury}}]{brune1992manipulation}%
  \BibitemOpen
  \bibfield  {author} {\bibinfo {author} {\bibfnamefont {M.}~\bibnamefont
  {Brune}}, \bibinfo {author} {\bibfnamefont {S.}~\bibnamefont {Haroche}},
  \bibinfo {author} {\bibfnamefont {J.~M.}\ \bibnamefont {Raimond}}, \bibinfo
  {author} {\bibfnamefont {L.}~\bibnamefont {Davidovich}}, \ and\ \bibinfo
  {author} {\bibfnamefont {N.}~\bibnamefont {Zagury}},\ }\href {\doibase
  10.1103/PhysRevA.45.5193} {\bibfield  {journal} {\bibinfo  {journal} {Phys.
  Rev. A}\ }\textbf {\bibinfo {volume} {45}},\ \bibinfo {pages} {5193}
  (\bibinfo {year} {1992})}\BibitemShut {NoStop}%
\bibitem [{\citenamefont {Vlastakis}\ \emph {et~al.}(2015)\citenamefont
  {Vlastakis}, \citenamefont {Petrenko}, \citenamefont {Ofek}, \citenamefont
  {Sun}, \citenamefont {Leghtas}, \citenamefont {Sliwa}, \citenamefont {Liu},
  \citenamefont {Hatridge}, \citenamefont {Blumoff}, \citenamefont {Frunzio}
  \emph {et~al.}}]{vlastakis2015characterizing}%
  \BibitemOpen
  \bibfield  {author} {\bibinfo {author} {\bibfnamefont {B.}~\bibnamefont
  {Vlastakis}}, \bibinfo {author} {\bibfnamefont {A.}~\bibnamefont {Petrenko}},
  \bibinfo {author} {\bibfnamefont {N.}~\bibnamefont {Ofek}}, \bibinfo {author}
  {\bibfnamefont {L.}~\bibnamefont {Sun}}, \bibinfo {author} {\bibfnamefont
  {Z.}~\bibnamefont {Leghtas}}, \bibinfo {author} {\bibfnamefont
  {K.}~\bibnamefont {Sliwa}}, \bibinfo {author} {\bibfnamefont
  {Y.}~\bibnamefont {Liu}}, \bibinfo {author} {\bibfnamefont {M.}~\bibnamefont
  {Hatridge}}, \bibinfo {author} {\bibfnamefont {J.}~\bibnamefont {Blumoff}},
  \bibinfo {author} {\bibfnamefont {L.}~\bibnamefont {Frunzio}},  \emph
  {et~al.},\ }\href@noop {} {\bibfield  {journal} {\bibinfo  {journal} {Nature
  communications}\ }\textbf {\bibinfo {volume} {6}},\ \bibinfo {pages} {8970}
  (\bibinfo {year} {2015})}\BibitemShut {NoStop}%
\bibitem [{\citenamefont {Kurizki}\ \emph {et~al.}(2015)\citenamefont
  {Kurizki}, \citenamefont {Bertet}, \citenamefont {Kubo}, \citenamefont
  {M{\o}lmer}, \citenamefont {Petrosyan}, \citenamefont {Rabl},\ and\
  \citenamefont {Schmiedmayer}}]{kurizki2015quantum}%
  \BibitemOpen
  \bibfield  {author} {\bibinfo {author} {\bibfnamefont {G.}~\bibnamefont
  {Kurizki}}, \bibinfo {author} {\bibfnamefont {P.}~\bibnamefont {Bertet}},
  \bibinfo {author} {\bibfnamefont {Y.}~\bibnamefont {Kubo}}, \bibinfo {author}
  {\bibfnamefont {K.}~\bibnamefont {M{\o}lmer}}, \bibinfo {author}
  {\bibfnamefont {D.}~\bibnamefont {Petrosyan}}, \bibinfo {author}
  {\bibfnamefont {P.}~\bibnamefont {Rabl}}, \ and\ \bibinfo {author}
  {\bibfnamefont {J.}~\bibnamefont {Schmiedmayer}},\ }\href {\doibase
  10.1073/pnas.1419326112} {\bibfield  {journal} {\bibinfo  {journal}
  {Proceedings of the National Academy of Sciences}\ }\textbf {\bibinfo
  {volume} {112}},\ \bibinfo {pages} {3866} (\bibinfo {year}
  {2015})}\BibitemShut {NoStop}%
\bibitem [{\citenamefont {Cochrane}\ \emph {et~al.}(1999)\citenamefont
  {Cochrane}, \citenamefont {Milburn},\ and\ \citenamefont
  {Munro}}]{cochrane1999macroscopically}%
  \BibitemOpen
  \bibfield  {author} {\bibinfo {author} {\bibfnamefont {P.~T.}\ \bibnamefont
  {Cochrane}}, \bibinfo {author} {\bibfnamefont {G.~J.}\ \bibnamefont
  {Milburn}}, \ and\ \bibinfo {author} {\bibfnamefont {W.~J.}\ \bibnamefont
  {Munro}},\ }\href {\doibase 10.1103/PhysRevA.59.2631} {\bibfield  {journal}
  {\bibinfo  {journal} {Phys. Rev. A}\ }\textbf {\bibinfo {volume} {59}},\
  \bibinfo {pages} {2631} (\bibinfo {year} {1999})}\BibitemShut {NoStop}%
\bibitem [{\citenamefont {Jeong}\ and\ \citenamefont
  {Kim}(2002)}]{jeong2002efficient}%
  \BibitemOpen
  \bibfield  {author} {\bibinfo {author} {\bibfnamefont {H.}~\bibnamefont
  {Jeong}}\ and\ \bibinfo {author} {\bibfnamefont {M.~S.}\ \bibnamefont
  {Kim}},\ }\href {\doibase 10.1103/PhysRevA.65.042305} {\bibfield  {journal}
  {\bibinfo  {journal} {Phys. Rev. A}\ }\textbf {\bibinfo {volume} {65}},\
  \bibinfo {pages} {042305} (\bibinfo {year} {2002})}\BibitemShut {NoStop}%
\bibitem [{\citenamefont {Leghtas}\ \emph {et~al.}(2013)\citenamefont
  {Leghtas}, \citenamefont {Kirchmair}, \citenamefont {Vlastakis},
  \citenamefont {Schoelkopf}, \citenamefont {Devoret},\ and\ \citenamefont
  {Mirrahimi}}]{leghtas2013hardware}%
  \BibitemOpen
  \bibfield  {author} {\bibinfo {author} {\bibfnamefont {Z.}~\bibnamefont
  {Leghtas}}, \bibinfo {author} {\bibfnamefont {G.}~\bibnamefont {Kirchmair}},
  \bibinfo {author} {\bibfnamefont {B.}~\bibnamefont {Vlastakis}}, \bibinfo
  {author} {\bibfnamefont {R.~J.}\ \bibnamefont {Schoelkopf}}, \bibinfo
  {author} {\bibfnamefont {M.~H.}\ \bibnamefont {Devoret}}, \ and\ \bibinfo
  {author} {\bibfnamefont {M.}~\bibnamefont {Mirrahimi}},\ }\href {\doibase
  10.1103/PhysRevLett.111.120501} {\bibfield  {journal} {\bibinfo  {journal}
  {Phys. Rev. Lett.}\ }\textbf {\bibinfo {volume} {111}},\ \bibinfo {pages}
  {120501} (\bibinfo {year} {2013})}\BibitemShut {NoStop}%
\bibitem [{\citenamefont {Andersen}\ \emph {et~al.}(2015)\citenamefont
  {Andersen}, \citenamefont {Neergaard-Nielsen}, \citenamefont {Van~Loock},\
  and\ \citenamefont {Furusawa}}]{andersen2015hybrid}%
  \BibitemOpen
  \bibfield  {author} {\bibinfo {author} {\bibfnamefont {U.~L.}\ \bibnamefont
  {Andersen}}, \bibinfo {author} {\bibfnamefont {J.~S.}\ \bibnamefont
  {Neergaard-Nielsen}}, \bibinfo {author} {\bibfnamefont {P.}~\bibnamefont
  {Van~Loock}}, \ and\ \bibinfo {author} {\bibfnamefont {A.}~\bibnamefont
  {Furusawa}},\ }\href@noop {} {\bibfield  {journal} {\bibinfo  {journal}
  {Nature Physics}\ }\textbf {\bibinfo {volume} {11}},\ \bibinfo {pages} {713}
  (\bibinfo {year} {2015})}\BibitemShut {NoStop}%
\bibitem [{\citenamefont {Penrose}(1996)}]{penrose1996gravity}%
  \BibitemOpen
  \bibfield  {author} {\bibinfo {author} {\bibfnamefont {R.}~\bibnamefont
  {Penrose}},\ }\href@noop {} {\bibfield  {journal} {\bibinfo  {journal} {Gen.
  Relativ. Gravit.}\ }\textbf {\bibinfo {volume} {28}},\ \bibinfo {pages} {581}
  (\bibinfo {year} {1996})}\BibitemShut {NoStop}%
\bibitem [{\citenamefont {Kleckner}\ \emph {et~al.}(2008)\citenamefont
  {Kleckner}, \citenamefont {Pikovski}, \citenamefont {Jeffrey}, \citenamefont
  {Ament}, \citenamefont {Eliel}, \citenamefont {Van Den~Brink},\ and\
  \citenamefont {Bouwmeester}}]{kleckner2008creating}%
  \BibitemOpen
  \bibfield  {author} {\bibinfo {author} {\bibfnamefont {D.}~\bibnamefont
  {Kleckner}}, \bibinfo {author} {\bibfnamefont {I.}~\bibnamefont {Pikovski}},
  \bibinfo {author} {\bibfnamefont {E.}~\bibnamefont {Jeffrey}}, \bibinfo
  {author} {\bibfnamefont {L.}~\bibnamefont {Ament}}, \bibinfo {author}
  {\bibfnamefont {E.}~\bibnamefont {Eliel}}, \bibinfo {author} {\bibfnamefont
  {J.}~\bibnamefont {Van Den~Brink}}, \ and\ \bibinfo {author} {\bibfnamefont
  {D.}~\bibnamefont {Bouwmeester}},\ }\href@noop {} {\bibfield  {journal}
  {\bibinfo  {journal} {New Journal of Physics}\ }\textbf {\bibinfo {volume}
  {10}},\ \bibinfo {pages} {095020} (\bibinfo {year} {2008})}\BibitemShut
  {NoStop}%
\bibitem [{\citenamefont {Gottesman}\ \emph {et~al.}(2001)\citenamefont
  {Gottesman}, \citenamefont {Kitaev},\ and\ \citenamefont
  {Preskill}}]{gottesman2001encoding}%
  \BibitemOpen
  \bibfield  {author} {\bibinfo {author} {\bibfnamefont {D.}~\bibnamefont
  {Gottesman}}, \bibinfo {author} {\bibfnamefont {A.}~\bibnamefont {Kitaev}}, \
  and\ \bibinfo {author} {\bibfnamefont {J.}~\bibnamefont {Preskill}},\
  }\href@noop {} {\bibfield  {journal} {\bibinfo  {journal} {Physical Review
  A}\ }\textbf {\bibinfo {volume} {64}},\ \bibinfo {pages} {012310} (\bibinfo
  {year} {2001})}\BibitemShut {NoStop}%
\bibitem [{\citenamefont {Weigand}\ and\ \citenamefont
  {Terhal}(2020)}]{weigand2019realizing}%
  \BibitemOpen
  \bibfield  {author} {\bibinfo {author} {\bibfnamefont {D.~J.}\ \bibnamefont
  {Weigand}}\ and\ \bibinfo {author} {\bibfnamefont {B.~M.}\ \bibnamefont
  {Terhal}},\ }\href {\doibase 10.1103/PhysRevA.101.053840} {\bibfield
  {journal} {\bibinfo  {journal} {Phys. Rev. A}\ }\textbf {\bibinfo {volume}
  {101}},\ \bibinfo {pages} {053840} (\bibinfo {year} {2020})}\BibitemShut
  {NoStop}%
\end{thebibliography}
\end{document}